\documentclass[11pt]{article}% добавляется twoside для двусторонней печати

\usepackage[T1]{fontenc}
\usepackage[cp1251]{inputenc}
\usepackage{textcomp}
\usepackage[centertags]{amsmath}
\usepackage[mediummath]{nccmath}
\usepackage{amsfonts}
\usepackage{amssymb}
\usepackage[pdftex]{hyperref}
\usepackage{graphicx}
\usepackage{graphbox}
\usepackage[numbers,sort&compress]{natbib}% упорядочивает ссылки

\usepackage{paperinitial}% Установка параметров страницы

\paperinitialization{15mm}{15mm}{15mm}{15mm}{2pt}{10pt}

\DeclareMathOperator{\re}{Re}
\DeclareMathOperator{\im}{Im}

\DeclareMathOperator{\rot}{rot}
\DeclareMathOperator{\diag}{diag}

\newcommand{\lan}{\langle}
\newcommand{\ran}{\rangle}
\newcommand{\bs}{\boldsymbol}

\newcommand{\e}{\varepsilon}
\newcommand{\vf}{\varphi}
\newcommand{\vk}{\varkappa}
\newcommand{\s}{\sigma}
\newcommand{\vs}{\varsigma}

\newcommand{\al}{\alpha}
\newcommand{\be}{\beta}
\newcommand{\ga}{\gamma}

\newcommand{\de}{\delta}
\newcommand{\De}{\Delta}

\newcommand{\la}{\lambda}

\newcommand{\ups}{\upsilon}

\newcommand{\spx}{\mathbf{x}}
\newcommand{\spy}{\mathbf{y}}

\newcommand{\spk}{\mathbf{k}}
\newcommand{\spe}{\mathbf{e}}
\newcommand{\R}{\mathbb{R}}

\begin{document}%\selectlanguage{english}
\allowdisplaybreaks[4]% позволяет переносить многострочные формулы
\frenchspacing% уменьшение пробелов после запятых
\setlength{\unitlength}{1pt}% устанавливает единицу длины в окружении picture

\title{{\Large\textbf{Radiation of twisted photons from charged particles moving in cholesterics}}}

\date{}

\author{
O.V. Bogdanov${}^{1),2)}$\thanks{E-mail: \texttt{bov@tpu.ru}},\;
P.O. Kazinski${}^{1)}$\thanks{E-mail: \texttt{kpo@phys.tsu.ru}},\;
P.S. Korolev${}^{1)}$\thanks{E-mail: \texttt{kizorph.d@gmail.com}},\;
and G.Yu. Lazarenko${}^{1)}$\thanks{E-mail: \texttt{laz@phys.tsu.ru}}\\[0.5em]
{\normalsize ${}^{1)}$ Physics Faculty, Tomsk State University, Tomsk 634050, Russia}\\
{\normalsize ${}^{2)}$ Tomsk Polytechnic University, Tomsk 634050, Russia}
}

\maketitle

\begin{abstract}

The radiation of twisted photons by charged particles traversing a cholesteric plate is studied in the framework of quantum electrodynamics in an anisotropic inhomogeneous dispersive medium. The complete set of solutions to the Maxwell equations in the cholesteric plate is constructed in the paraxial and small anisotropy approximations. The explicit expressions for the average numbers of plane-wave and twisted photons created by a charged point particle crossing the cholesteric plate are derived. The selection rules for the twisted photons radiated at the harmonic $n\in \mathbb{Z}$ are established. In the paraxial regime, the projection of the orbital angular momentum of a radiated twisted photon obeys the selection rule $l=\pm(2n+1)$. In the approximation of a small anisotropy of the permittivity tensor, the selection rule becomes $m=\pm2n$, where $m$ is the projection of the total angular momentum of a radiated twisted photon. The sign ``$\pm$'' in these selection rules is related to the choice of the forward or reflected waves in the cholesteric and is realized at the different energies of radiated photons. As the examples, the radiation of optical twisted photons by electrons with the Lorentz factors $\ga=235$ and $\ga=500$ and by uranium nuclei with $\ga=2$ are considered. It is shown that charged particles traversing normally a cholesteric plate can be used as a pure source of twisted photons.

\end{abstract}

\section{Introduction}

Transition radiation of relativistic particles in periodic dispersive media is a well studied subject both theoretically and experimentally \cite{Ginzburg,BazylZhev,GinzbThPhAstr}. It was proved in the recent paper \cite{BKL5} that the media possessing a helical symmetry can be used for generation of twisted photons \cite{MolTerTorTor,JenSerprl,JenSerepj,PadgOAM25,Roadmap16,KnyzSerb}. Namely, the radiation of charged particles traversing such media along the axis of the helical symmetry is a pure source of twisted photons. In the present paper, we investigate a particular case of such a media -- a cholesteric liquid crystal (CLC) -- and show by explicit calculations that the charged particle crossing the CLC layer creates the photons with definite projection of the angular momentum at a given harmonic.

The transition radiation of plane-wave photons from charged particles moving in CLCs is rather well studied in the paraxial \cite{Ship91} and small anisotropy \cite{BelOrl72,ShipBel78,BelDmitOrl,Belyakov86,VelRey17,VelReyVaz17} approximations. However, the production of twisted photons in such systems was not investigated. An exception, in a certain sense, is the paper \cite{VoloLavr} where the twisted photons were created by transmitting the electromagnetic wave with a plane wavefront through the CLC layer with a dislocation. Other ways of generation of twisted photons by liquid crystals can be found, for example, in \cite{Barboza15}. In the present paper, we thoroughly investigate the generation of twisted photons by charged particles moving in CLCs and, in particular, construct the complete set of solutions to the Maxwell equations for the CLC plate. The latter allows one to find the radiation from charged particles moving along arbitrary trajectories and not just along a straight line.

Unfortunately, the works \cite{Ship91,BelDmitOrl,Kats,Chandra77,Blinov78,BelSon82,Osadch84,Osadch85,deGennProst,AksValRom04,AksKryuRom06,YangWu06,ShipPolRub,VetTimShab20} do not contain the complete set of solutions to the Maxwell equations for the CLC plate of a finite width in the form that can be immediately employed to attack the problem we study. That is why we reconsider this issue and, in addition, construct quantum electrodynamics (QED) in such a medium. Even for the simplest model of CLC permittivity \cite{LandLifshECM,YangWu06,deGennProst,Andri,BelSon82,Blinov78,Chandra77,VetTimShab20,BelDmitOrl}, the Maxwell equations are not exactly solvable. Therefore, in order to construct the complete set of solutions to the Maxwell equations, we use the paraxial and the small anisotropy approximations. These approximations allow one to describe the created radiation in the different regions of the parameter space. We suppose that the interfaces of the CLC plate are orthogonal to the CLC axis and the width of the plate equals an integer number of the permittivity periods. In that case it is not difficult to join the solutions of the Maxwell equations in the CLC and in vacuum. However, the joining coefficients turn out to be rather bulky. The complete set of solutions is used to construct the quantum electromagnetic field operator and QED in this medium. After that, the average number of photons created by classical currents and the inclusive probability to record such a photon are readily found \cite{Glaub51,Glaub2,KlauSud,BKL2,ippccb,GinzbThPhAstr}.

As a result, we obtain the explicit expressions for the average numbers of plane-wave and twisted photons radiated by a charged particle moving uniformly and rectilinearly and traversing a CLC plate. It follows from these expressions that such a system is a pure source of twisted photons in the sense that only the twisted photons with definite projection of the angular momentum are radiated at the $n$-th harmonic. In the paraxial approximation, the selection rule looks like $l=\pm(2n+1)$, where $l$ is the projection of the orbital angular momentum of the twisted photon, whereas in the small anisotropy approximation $m=\pm2n$, where $m$ is the projection of the total angular momentum of the twisted photon. The choice of the plus sign or minus sign in the selection rules depends on the radiation mode considered: the plus sign corresponds to the forward wave, whereas the minus sign is for the reflected one. These opposite signs are realized at the different energies of radiated twisted photons. Notice that these selection rules are in accordance with the general selection rule proved in \cite{BKL5} for helical media.

The paper is organized as follows. In Sec. \ref{Gener_Form}, QED in an anisotropic inhomogeneous dispersive transparent medium is constructed. In Sec. \eqref{CholLiqCrys}, the general formulas for the Maxwell equations in CLCs and for the twisted photons are collected. In particular, it is shown that the Maxwell equations in CLCs are equivalent to a one-dimensional matrix Schr\"{o}dinger equation. Section \ref{ParaxAppr} is devoted to the paraxial approximation to the solution of the Maxwell equations. In Sec. \ref{ModeFuncPar}, we obtain the complete set of solutions in the CLC plate in this approximation, join them with the solutions in vacuum, and construct the quantum operator of the electromagnetic field. Then, in Sec. \ref{RadProbSec}, we derive the explicit expressions for the average number of plane-wave and twisted photons produced by a charged particle that moves uniformly and rectilinearly and crosses the CLC plate. In Sec. \ref{AnisotrAppr}, the perturbation theory with respect to anisotropy of the permittivity tensor is developed. In this approximation, the mode functions of a quantum electromagnetic field and the average numbers of radiated plane-wave and twisted photons are deduced there. In Conclusion, the results of the paper are summarized. In Appendix \ref{UnitRels}, the unitarity relations for the joining coefficients are derived. The explicit expressions for these coefficients and the normalization constants are presented in Appendix \ref{JoinModFun}.

We borrow the notation from \cite{BKL2}. In particular, the system of units is used such that $\hbar=c=1$ and $e^2=4\pi\al$, where $\al$ is the fine structure constant.

\section{General formulas}\label{Gener_Form}

Let us consider the quantization of an electromagnetic field interacting with a classical current in an anisotropic inhomogeneous dispersive medium. The quantization procedure developed below is a generalization to anisotropic media of the analogous procedure given in \cite{BKL5} (see also \cite{Ginz40,Sokol40,SokolLosk,Riazanov57,Riazanov58,SokolLosk1,AbrGorDzyal,AleksNik}). We suppose that the spatial dispersion of the medium is negligibly small and the magnetic permeability $\mu_{ij}(k_0)=1$, where $k_0$ is the frequency of photons in the electromagnetic wave. The permittivity tensor satisfies \cite{LandLifshECM}
\begin{equation}\label{transpar_cond}
    \e_{ij}(k_0)=\e_{ji}^*(k_0),\quad k_0\in \mathbb{R},
\end{equation}
i.e., we assume that the medium is transparent. The medium transparency is needed for unitarity of QED we are going to construct. We also assume that $\e_{ij}(k_0)$ are analytic on some segment of the real axis. Then the transparency condition \eqref{transpar_cond} gives rise to
\begin{equation}\label{transpar_cond1}
    \e_{ij}(k_0)=\e_{ji}^*(k^*_0),\quad k_0\in \mathbb{C}.
\end{equation}
The real-valuedness of the electromagnetic fields implies
\begin{equation}\label{realvaluedness}
    \e_{ij}^*(k_0)=\e_{ij}(-k^*_0).
\end{equation}
The dynamics of the electromagnetic fields in a medium are causal when the functions $\e_{ij}(iz)$ are analytic functions of $z$ without singularities for $\re z>0$. Then it follows from the transparency condition \eqref{transpar_cond1} that the singularities of  $\e_{ij}(k_0)$ can be positioned on the real axis only. We additionally suppose that these singularities are isolated.

The free Maxwell equations for the Fourier transform of the electromagnetic potential in the generalized Coulomb gauge have the form \cite{LandLifshECM,RyazanovB,AbrGorDzyal}
\begin{equation}\label{Max_eqns}
    H_{ij}(k_0)A_j(k_0,\spx)=0,\qquad\partial_i(\e_{ij}(k_0,\spx)A_j(k_0,\spx))=0,
\end{equation}
where
\begin{equation}
    H_{ij}(k_0):=k_0^2\e_{ij}(k_0,\spx)-\rot^2_{ij}.
\end{equation}
The second equation in \eqref{Max_eqns} is the generalized Coulomb gauge. It ensues from the first equation under the assumption that $k_0\neq0$. Hereinafter we suppose that the electromagnetic field obeys such boundary conditions that the first equation in \eqref{Max_eqns} does not possess solutions for $k_0=0$.

In order to construct the operator of a quantum electromagnetic field, one has to find the complete set of solutions to \eqref{Max_eqns} that we will denote as $\psi_\al(k_{0\al},\spx)$, where $\al$ numbers the solutions and $k_0=k_{0\al}$. It is useful to consider Eq. \eqref{Max_eqns} out of the mass-shell \cite{FursVass,KalKaz2.12}
\begin{equation}\label{Max_offshell}
    H_{ij}(k_0)\psi_j(k_0,\spx)=\la(k_0)\psi_i(k_0,\spx),\quad\la(k_0)=const\in \mathbb{C}.
\end{equation}
On imposing the proper boundary conditions and setting $k_0\in \R$, the operator on the left-hand side of \eqref{Max_offshell} is self-adjoint on the space of square-integrable complex vector-functions with the scalar product
\begin{equation}
    \lan \psi,\vf\ran:=\int d\spx\psi^*_i(\spx)\vf_i(\spx).
\end{equation}
At every fixed $k_0\in\R$ out of the singular points of $\e_{ij}(k_0)$, we have the complete set of solutions of \eqref{Max_offshell}:
\begin{equation}\label{Max_offshell1}
   H_{ij}(k_0)\psi_{\al j}(k_0,\spx)=\la_\al(k_0)\psi_{\al i}(k_0,\spx),\quad\la_\al(k_0)\in \R.
\end{equation}
For simplicity, we suppose that the spectrum is discrete, i.e., $\al$ runs a discrete set. This can always be achieved by placing the system in a sufficiently large box. Moreover, we assume that the spectrum $\la_\al(k_0)$ is nondegenerate which is valid for $\e_{ij}(k_0,\spx)$ taken in a general position.

The symmetry property \eqref{realvaluedness} and Eq. \eqref{Max_offshell1} taken at $k_0\in \mathbb{C}$ imply that
\begin{equation}
    H_{ij}(k_0)\psi^*_{\al j}(-k^*_0,\spx)=\la^*_\al(-k^*_0)\psi^*_{\al i}(-k^*_0,\spx).
\end{equation}
Therefore, there exists $\al'(\al)$ for any $\al$ such that $\la^*_{\al'(\al)}(-k^*_0)=\la_\al(k_0)$ and
\begin{equation}
    \psi^*_{\al i}(-k_0^*,\spx)=\psi_{\al'(\al) i}(k_0,\spx).
\end{equation}
It is clear that
\begin{equation}
    \al'(\al'(\al))=\al.
\end{equation}
Since $\e_{ij}(k_0)$  possess only isolated singularities on the real axis and the spectrum $\la_\al(k_0)$ is discrete, the functions $\psi_{\al i}(k_0,\spx)$ and $\la_\al(k_0)$ are analytic with respect to $k_0$ near the real axis \cite{ReedSim4}. Besides,
\begin{equation}\label{symm_lambda}
    \la^*_\al(k_0)=\la_\al(k^*_0).
\end{equation}
In fact, this symmetry relation is a consequence of the transparency condition \eqref{transpar_cond}. The mass-shell condition becomes
\begin{equation}\label{mass-shell}
    \la_\al(k_0)=0.
\end{equation}
If $k_{0\al}$ is the solution to this equation, then $\la_{\al'(\al)}(-k_{0\al})=0$. In other words, the allowable frequencies always come in pairs $k_{0\al}$ and $-k_{0\al}$. Taking the divergence of \eqref{Max_offshell1}, one obtains that the mode functions $\psi_{\al i}(k_{0\al},\spx)$ restricted to the mass-shell obey the generalized Coulomb gauge \eqref{Max_eqns}.

The retarded Green's function for the operator \eqref{Max_eqns} is written as
\begin{equation}\label{ret_Gr_func}
    G^{-}_{ij}(t;t')=\int_C\frac{dk_0}{2\pi}H_{ij}^{-1}(k_0)e^{-ik_0(t-t')},
\end{equation}
where the contour $C$ goes from left to right a little bit higher than the real axis. The requirement of causality of solutions of the Maxwell equations \eqref{Max_eqns} brings about the operator $H_{ij}^{-1}(k_0)$ to be analytic for $\im k_0>0$. Therefore, $\la_\al(k_0)$ do not have zeros for $\im k_0>0$. The symmetry \eqref{symm_lambda} implies that the operator $H_{ij}^{-1}(k_0)$ does not possess the singularities for $\im k_0<0$ as well. All the singularities of $H_{ij}^{-1}(k_0)$ lie on the real axis and correspond to the spectrum $\la_\al(k_0)$ taken on the mass-shell \eqref{mass-shell}. The advanced Green's function $G^+_{ij}(t;t')$ has the form \eqref{ret_Gr_func}, where the contour $C$ runs from left to right a little bit lower that the real axis. The commutator Green's function is
\begin{equation}
    \tilde{G}_{ij}(t,t')=i[G^-_{ij}(t,t') -G^+_{ij}(t,t')]=i\int_{\tilde{C}}\frac{dk_0}{2\pi}H_{ij}^{-1}(k_0)e^{-ik_0(t-t')},
\end{equation}
where the contour $\tilde{C}$ passes clockwise around the real axis. Evaluating this integral by residues, we obtain
\begin{equation}\label{comm_Gr_func}
    \tilde{G}_{ij}(t,\spx;t',\spy)=\sum_\be \Big[f_\be \psi_{\be i}(k_{0\be},\spx)\psi^*_{\be j}(k_{0\be},\spy) e^{-ik_{0\be}(t-t')}-c.c.\Big],\qquad f^{-1}_\be:=\la'_\be(k_{0\be}),
\end{equation}
where the symmetry property of the eigenvalues $\la_\al(k_0)$ was used, it is supposed that $\la'_\al(k_{0\al})\neq0$, and the sum over $\beta$ is carried over those eigenvalues that satisfy $\la'_\be(k_{0\be})>0$. The last inequality defines the splitting of the modes into positive- and negative-frequency ones \cite{MigdalMM,FursVass,KalKaz2.12,ippccb}.

On the other hand, let us introduce the quantum electromagnetic field operator in the interaction representation
\begin{equation}\label{quant_field}
    \hat{A}_i(t,\spx):=\sum_\be\Big[\hat{a}_\be f^{1/2}_\be \psi_{\be i}(k_{0\be};\spx) e^{-ik_{0\be}t} +\hat{a}^\dag_\be f^{1/2}_\be \psi^*_{\be i}(k_{0\be};\spx)e^{ik_{0\be}t} \Big],
\end{equation}
where $\hat{a}_\be$, $\hat{a}^\dag_\be$ are the creation-annihilation operators
\begin{equation}
    [\hat{a}_\be,\hat{a}^\dag_{\be'}]=\de_{\be\be'},\qquad [\hat{a}_\be,\hat{a}_{\be'}]=0.
\end{equation}
This quantum field satisfies Eqs. \eqref{Max_eqns}. Its commutator Green's function,
\begin{equation}
    \tilde{G}_{ij}(t,\spx;t',\spy)=[\hat{A}_i(t,\spx),\hat{A}_j(t',\spy)],
\end{equation}
coincides with \eqref{comm_Gr_func}.

The construction of quantum electrodynamics in an anisotropic inhomogeneous dispersive medium is further developed as in Secs. 2, 3 of \cite{BKL5}. In particular, the mode functions of the quantum field taken at the discontinuities of the permittivity tensor should satisfy the boundary conditions \cite{LandLifshECM,RyazanovB,AbrGorDzyal}
\begin{equation}\label{bound_conds}
    [\psi_\tau]=0,\qquad[(\rot\psi)_\tau]=0,
\end{equation}
where $\tau$ denotes the vector components tangent to the discontinuity surface and the square brackets denote the jump of the corresponding quantity on this surface. The boundary conditions on the surface of an ideal conductor are as follows
\begin{equation}\label{bound_conds_cond}
    \psi_\tau=0,
\end{equation}
and $\psi=0$ inside of the conductor. Furthermore, the formulas (50), (52), and (53) of \cite{BKL5} for the average number of photons created by the current and for the inclusive probability to detect a photon are valid (see also \cite{ippccb}).

%\newpage
\section{Cholesteric liquid crystals}\label{CholLiqCrys}

The simplest model of a cholesteric is described by the permittivity tensor of the form \cite{LandLifshECM,YangWu06,deGennProst,BelSon82,Blinov78,Chandra77,Andri}
\begin{equation}\label{permit_holec}
    \e_{ij}(k_0)=\e_\perp(k_0)\de_{ij}+(\e_\parallel(k_0)-\e_\perp(k_0))\tau_i\tau_j,
\end{equation}
where the director is
\begin{equation}
    \tau_i=(\cos(qz),\sin(qz),0),
\end{equation}
and $q$ characterizes the period of the director variation. As a rule, the quantity
\begin{equation}
    \De\e(k_0):=\e_\parallel(k_0)-\e_\perp(k_0),
\end{equation}
is small and can take positive or negative values. We suppose that the CLC constitutes the plate with the width $L=\pi N_u/|q|$, where $N_u$ is the number of periods. This plate is placed in vacuum and is perpendicular to the $z$ axis (the axis $3$).

In the previous section we saw that, in order to construct the quantum field \eqref{quant_field}, one has to find the complete set of solutions of the Maxwell equations \eqref{Max_eqns}. Such a procedure for the plate made of a homogeneous isotropic dielectric was realized in Sec. 5.A of \cite{BKL5} in terms of twisted photons. As for the CLC plate, the procedure is analogous. In particular, the general formula (92) of \cite{BKL5} for the average number of photons produced by the classical current is valid and the formula (52) of \cite{BKL5} for the inclusive probability to record a photon holds. The complete set of solution and transition radiation of plane-wave photons for the plate made of a homogeneous isotropic dielectric are described in \cite{Pafomov}.

In the case at hand, it is convenient to construct a complete set of solutions in terms of plane-wave photons and then, taking their linear combination as in \cite{JenSerprl,JenSerepj}, to obtain the twisted photons. Introducing the basis $\{\spe_+,\spe_-,\spe_3\}$, where
\begin{equation}
    \spe_\pm:=\spe_1\pm i \spe_2,
\end{equation}
any vector $\mathbf{A}$ can be decomposed as
\begin{equation}\label{cylin_basis}
    \mathbf{A}=\frac{1}{2}(\spe_+ A_-+\spe_- A_+)+\spe_3 A_3.
\end{equation}
The expansion of a plane wave in terms of twisted ones has the form
\begin{equation}\label{pln_waves_to_twisted}
    \frac{\mathbf{f}(s,\spk)e^{i\spk\spx}}{\sqrt{2k_0V}}=-\frac{\sin\theta e^{ik_3z}}{2\sqrt{k_0V}} \sum_{m=-\infty}^\infty i^me^{-im\vf}\sqrt{RL_z} \Big(\frac{2}{\sin\theta}\Big)^{3/2} \bs{\Phi}(s,m,k_3,k_\perp;\spx),
\end{equation}
where the polarization vector of a plane-wave photon with helicity $s$ in the Cartesian basis reads
\begin{equation}
    \mathbf{f}(s,\spk)=(\cos\vf\cos\theta-is\sin\vf,\sin\vf\cos\theta+is\cos\vf,-\sin\theta)/\sqrt{2},
\end{equation}
and $k_\perp=|k_+|$, $\vf=\arg k_+$, $\sin\theta:=k_\perp/k_0\equiv n_\perp$, and $n_3=\cos\theta$. The components of the mode functions of twisted photons are written as \cite{GottfYan,JaurHac,BiaBirBiaBir,JenSerprl,JenSerepj}
\begin{equation}
\begin{split}
    \Phi_3(s,m,k_3,k_\perp;\spx)&=\frac{1}{\sqrt{RL_z}} \Big(\frac{\sin\theta}{2}\Big)^{3/2}j_m(k_\perp x_+,k_\perp x_-), \\ \Phi_\pm(s,m,k_3,k_\perp;\spx)&=\frac{i\sin\theta}{s\pm\cos\theta}\Phi_3(s,m\pm1,k_3,k_\perp;\spx),\\
    j_m(k_\perp x_+,k_\perp x_-)&=J_m(k_\perp|x_+|)e^{im\arg x_+}.
\end{split}
\end{equation}
The parameters $V$, $R$, and $L_z$ characterize the normalization volume.

We suppose that the plate is located at $z\in[-L,0]$ and the detector is placed in the region $z>0$. Therefore, we have a plane wave in the domain $z>0$ that is joined at $z=0$ with the solution of \eqref{Max_eqns} in the CLC that in turn is joined at $z=-L$ with the linear combination of plane waves propagating in the region $z<-L$. The joining conditions \eqref{bound_conds} become
\begin{equation}\label{bound_conds_chol}
    [A_\pm]_{z=0}=[A_\pm]_{z=-L}=0,\qquad [\rot A_\pm]_{z=0}=[\rot A_\pm]_{z=-L}=0.
\end{equation}
Notice that
\begin{equation}
    \rot A_\pm=\pm i(\partial_3A_\pm -\partial_\pm A_3),\qquad \rot A_3=\frac{i}{2}(\partial_+ A_- -\partial_-A_+).
\end{equation}
In Secs. \ref{ParaxAppr}, \ref{AnisotrAppr}, we shall use these conditions to construct the mode functions of a quantum electromagnetic field in the whole space. 

Let us find the complete set of solutions of the Maxwell equations in the CLC plate. We seek for the solution in the form
\begin{equation}\label{pln_waves}
    A_i(k_0,\spx)=e^{i\spk_\perp\spx_\perp}A_i(k_0,z).
\end{equation}
Substituting \eqref{pln_waves}, \eqref{cylin_basis} into \eqref{Max_eqns}, we deduce
\begin{equation}\label{Max_eqns1}
\begin{split}
    \frac{\bar{k}_0^2}{2}\big[\de\e e^{2i qz}A_- +(2+\de\e)A_+\big] +\frac{k_+^2}{2}A_- +\Big(\partial_3^2-\frac{k_\perp^2}{2}\Big)A_+ -ik_+\partial_3A_3&=0,\\
    \frac{\bar{k}_0^2}{2}\big[\de\e e^{-2iqz}A_+ +(2+\de\e)A_-\big] +\frac{k_-^2}{2}A_+ +\Big(\partial_3^2-\frac{k_\perp^2}{2}\Big)A_- -ik_-\partial_3A_3&=0,\\
    (\bar{k}_0^2-k_\perp^2)A_3 -\frac{ik_+}{2}\partial_3A_- -\frac{ik_-}{2}\partial_3A_+&=0,
\end{split}
\end{equation}
where $\bar{k}_0^2:=k_0^2\e_\perp$ and $\de\e:=\De\e/\e_\perp$. It is useful to introduce $\bar\theta:=qz-\vf$ and
\begin{equation}\label{apm_not}
    a_\pm=A_\pm e^{\mp i\vf}.
\end{equation}
Then, taking $A_3$ from the last equation in the system \eqref{Max_eqns1},
\begin{equation}\label{a_3_comp}
    A_3=:a_3=\frac{ik_\perp}{2(\bar{k}_0^2-k_\perp^2)}\partial_3(a_++a_-),
\end{equation}
and substituting into the first two ones, we have
\begin{equation}\label{Max_eqns2}
    (\partial_3K\partial_3+V)
        \left[
      \begin{array}{c}
        a_+ \\
        a_- \\
      \end{array}
    \right]=0,
\end{equation}
where
\begin{equation}
    K=\left[
      \begin{array}{cc}
         1+\frac{k_\perp^2}{2\bar{k}_3^2} & \frac{k_\perp^2}{2\bar{k}_3^2} \\
         \frac{k_\perp^2}{2\bar{k}_3^2} & 1+\frac{k_\perp^2}{2\bar{k}_3^2} \\
      \end{array}
    \right],\qquad
    V=-
    \frac{k_\perp^2}{2}
    \left[
      \begin{array}{cc}
         1 & -1 \\
         -1 & 1 \\
      \end{array}
    \right]+
    \frac{\bar{k}_0^2}{2}
    \left[
      \begin{array}{cc}
        2+\de\e & \de\e e^{2i\bar{\theta}} \\
        \de\e e^{-2i\bar{\theta}} & 2+\de\e \\
      \end{array}
    \right],
\end{equation}
and $\bar{k}_3^2:=\bar{k}_0^2-k_\perp^2$. It is evident that the transformation
\begin{equation}\label{symm_transf}
    \left[
       \begin{array}{c}
         a_+(z) \\
         a_-(z) \\
       \end{array}
     \right]\rightarrow
     \left[
       \begin{array}{c}
         a^*_-(z) \\
         a^*_+(z) \\
       \end{array}
     \right]
\end{equation}
is a symmetry of Eq. \eqref{Max_eqns2}. Therefore, it is sufficient to find the two linear independent solutions of Eq. \eqref{Max_eqns2}.

The equation \eqref{Max_eqns2} can be cast into the form of a stationary matrix Schr\"{o}dinger equation. Introduce
\begin{equation}
    \left[
      \begin{array}{c}
        a_+ \\
        a_- \\
      \end{array}
    \right]=:K^{-1/2}
    \left[
      \begin{array}{c}
        \be_+ \\
        \be_- \\
      \end{array}
    \right],
\end{equation}
where
\begin{equation}
    K^{-1/2}=\frac{1}{2}
    \left[
      \begin{array}{cc}
        \tanh\eta+1 & \tanh\eta-1 \\
        \tanh\eta-1 & \tanh\eta+1 \\
      \end{array}
    \right],
\end{equation}
and $\sinh\eta:=\bar{k}_3/k_\perp$. Multiplying Eq. \eqref{Max_eqns2} by $K^{-1/2}$ from the left, we come to the matrix Schr\"{o}dinger equation
\begin{equation}\label{Max_eqns2-1}
    \bigg\{\partial_3^2-
    \frac{k_\perp^2}{2}
    \left[
      \begin{array}{cc}
        c & d \\
        d^* & c \\
      \end{array}
    \right]\bigg\}
    \left[
      \begin{array}{c}
        \be_+ \\
        \be_- \\
      \end{array}
    \right]=0,
\end{equation}
where
\begin{equation}
    c=2\sinh^2\eta +\de\e\sin(\bar{\theta}+i\eta)\sin(\bar{\theta}-i\eta) ,\qquad
    d=-\de\e\sin^2(\bar{\theta}-i\eta).
\end{equation}
The equation \eqref{Max_eqns2-1} is integrable for $\de\e=0$, $k_\perp=0$ ($\bar{k}_3\neq0$), and $\bar{k}_3=0$ ($k_\perp\neq0$).

We shall also need the other representation of Eq. \eqref{Max_eqns2}. If $N_u$ is large, the solution of \eqref{Max_eqns2} can be written in the form
\begin{equation}\label{Fourier_ser}
    a_\pm(z)=\int_0^{2q} \frac{L d\kappa_3}{2\pi} \sum_{l=-\infty}^\infty a_{\pm}(\kappa_3+2ql) e^{i(\kappa_3+2ql)(z-\vf/q)},
\end{equation}
where $\kappa_3\in[0,2q)$ is a quasi-momentum and $p_3:=\kappa_3+2ql$ is the physical momentum of the mode in the expansion \eqref{Fourier_ser}. The quasi-momentum $\kappa_3=2qn/N_u$, where $n=\overline{0,N_u-1}$. Substituting \eqref{Fourier_ser} into \eqref{Max_eqns2}, we arrive at the infinite system of coupled algebraic equations
\begin{equation}\label{Max_eqns3}
\begin{split}
    M(p_3)a_{+}(p_3)&=-\frac{\bar{k}_0^2}{2}\de\e\big[a_{+}(p_3) +a_{-}(p_3-2q)\big] +\frac{k_\perp^2}{4}\de\e\big[a_{+}(p_3)+a_{-}(p_3) +a_{+}(p_3+2q) +a_{-}(p_3-2q) \big],\\
    M(p_3)a_{-}(p_3)&=-\frac{\bar{k}_0^2}{2}\de\e\big[a_{-}(p_3) +a_{+}(p_3+2q)\big] +\frac{k_\perp^2}{4}\de\e\big[a_{+}(p_3)+a_{-}(p_3) +a_{+}(p_3+2q) +a_{-}(p_3-2q) \big],
\end{split}
\end{equation}
where
\begin{equation}
    M(p_3):=\bar{k}_3^2-p_3^2.
\end{equation}
The system of equations \eqref{Max_eqns3} fall into $N_u$ independent subsystems with the unknowns $a_{\pm}(\kappa_3+2ql)$, $l\in \mathbb{Z}$, where $\kappa_3$ numerates the subsystems. The solution of the systems \eqref{Max_eqns3} with the reversed sign of $q$ is obtained from the solution of the initial system by the transform
\begin{equation}
    q\rightarrow-q,\qquad a_{\pm}(\cdot)\rightarrow a_{\mp}(\cdot).
\end{equation}
Therefore, we further consider only the case of positive $q$, i.e., the right-handed CLCs. In the general case we have not succeeded in finding the exact solution of the system \eqref{Max_eqns3}. It is clear from \eqref{Max_eqns3} that this system is exactly solvable for $\de\e=0$ or $k_\perp=0$. Therefore, we shall consider below the perturbation theories with respect to $\de\e$ and $k_\perp$ allowing one to find the approximate solution of \eqref{Max_eqns3} or \eqref{Max_eqns2} in the different regions of the parameter space.

\section{Paraxial approximation}\label{ParaxAppr}
\subsection{Mode functions}\label{ModeFuncPar}

In this section, we consider the perturbative solution of Eq. \eqref{Max_eqns3} treating $k_\perp$ as a small parameter. As was already mentioned, it is sufficient to find the two linear independent solutions of Eq. \eqref{Max_eqns3}. In the leading order, at $k_\perp=0$, the system \eqref{Max_eqns3} possesses the solutions \cite{Ship91,BelDmitOrl,Kats,Chandra77,Blinov78,BelSon82,deGennProst,VetTimShab20,Mauguin11,deVries51}
\begin{equation}\label{sol_par}
\begin{split}
    a^{(1)}_+(p_3)=1,\qquad a^{(1)}_-(p_3-2q)=a^{(1)}_{0-}(p_3-2q) =\frac{2}{\de\e\bar{k}_0^2}\big[p_3^2-\bar{k}_0^2(1+\de\e/2)\big],\qquad p_3=p_0^{(1)}=q+v_+;\\
    a^{(2)}_+(p_3)=1,\qquad a^{(2)}_-(p_3-2q)=a^{(2)}_{0-}(p_3-2q)=\frac{2}{\de\e\bar{k}_0^2}\big[p_3^2-\bar{k}_0^2(1+\de\e/2)\big],\qquad p_3=p_0^{(2)}=q+v_-;
\end{split}
\end{equation}
where
\begin{equation}
    v_\pm:=\big[(1+\de\e/2)\bar{k}_0^2\pm 2q\bar{k}_0w+q^2\big]^{1/2},\qquad w:=\big[1+\de\e/2+\de\e^2\bar{k}_0^2/(16q^2)\big]^{1/2}.
\end{equation}
The rest two solutions are obtained with the aid of the symmetry transform \eqref{symm_transf}. It is assumed in \eqref{sol_par} that $v_-$ is real. Recall that we also suppose that $\e_\perp>0$ and $\e_\parallel>0$, and so $\de\e>-1$. The requirement of real-valuedness of $v_-$ implies that the interval of energies of the second mode, $\bar{k}_0\in |q|((1+\de\e)^{-1/2},1)$ for $\de\e>0$, and $\bar{k}_0\in |q|(1,(1+\de\e)^{-1/2})$ for $\de\e<0$, is forbidden. More explicitly, the forbidden energy band of the mode $2$ is written as
\begin{equation}
    k_0\in |q|(\e_\parallel^{-1/2},\e_\perp^{-1/2}),\quad\de\e>0;\qquad k_0\in |q|(\e_\perp^{-1/2},\e_\parallel^{-1/2}),\quad\de\e<0.
\end{equation}
Below we consider the energies of photons out of this range. A wave packet comprised of the modes $1$ moves to the left, whereas a wave packet consisting of the modes $2$ propagates to the right for the energies above the forbidden band and to the left for the energies below the forbidden band.

\begin{figure}[tp]
\centering
\includegraphics*[width=0.49\linewidth]{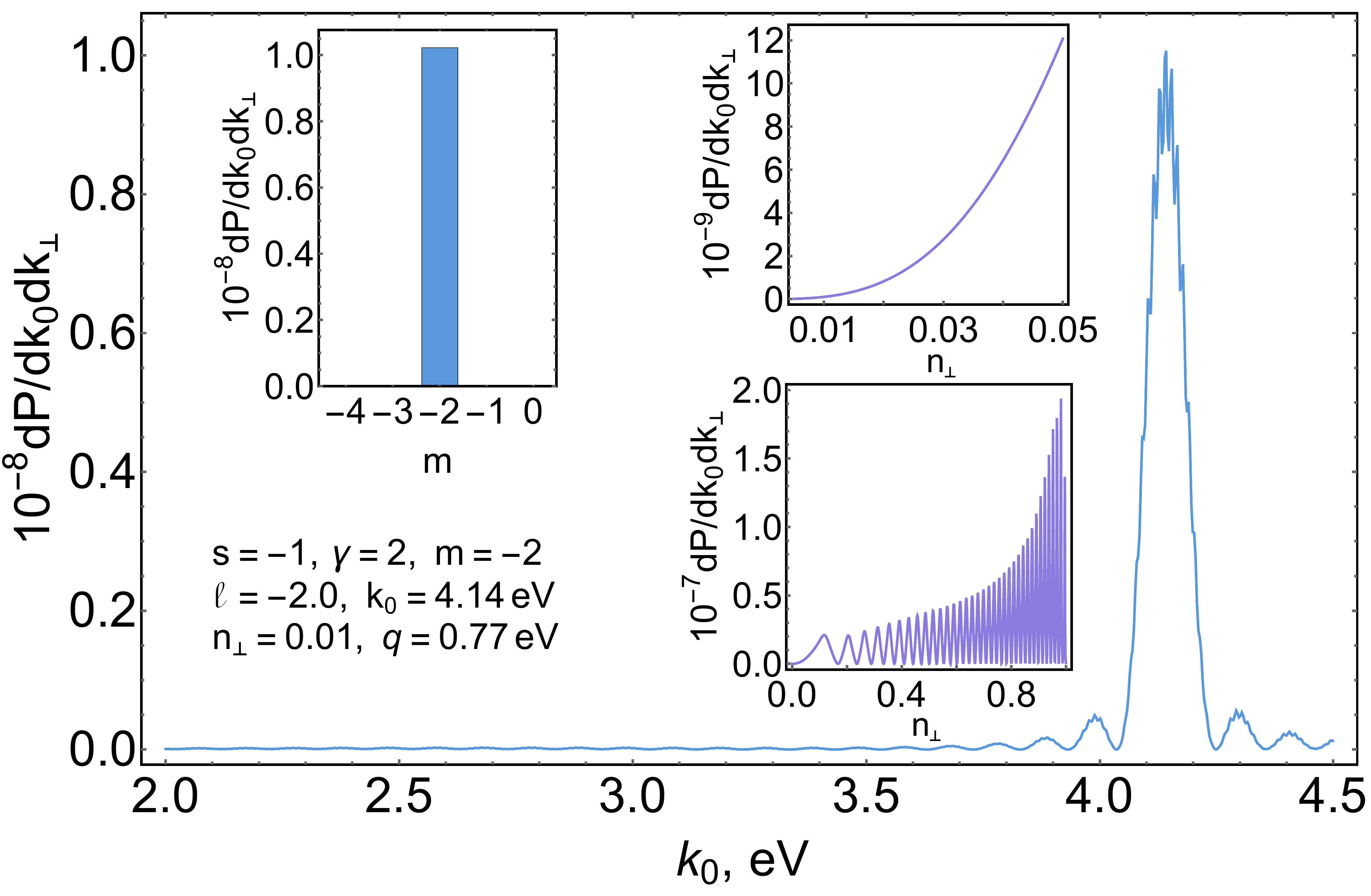}\,
\includegraphics*[width=0.49\linewidth]{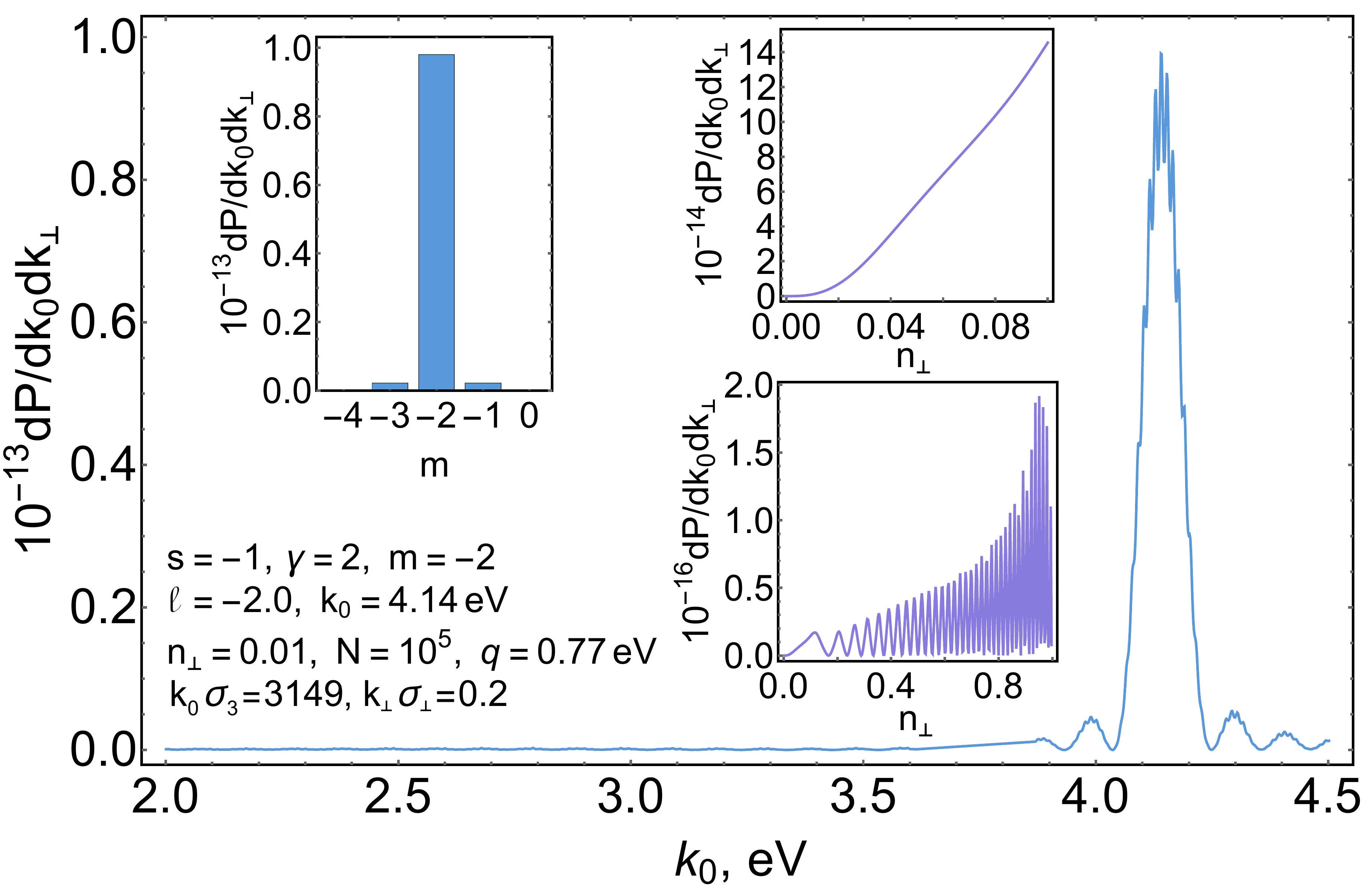}
\caption{{\footnotesize The average number of twisted photons produced by the ions ${}^{238}$U${}^{92+}$ with the Lorentz factor $\ga=2$ traversing normally the CLC plate of width $L=32$ $\mu$m with the number of periods $N_u=40$ corresponding to $q\approx0.77$ eV and the CLC helix pitch $1.6$ $\mu$m. The parameters of the permittivity tensor are taken to be $\e_\perp=2.22$ and $\e_\parallel=2.49$ (see, e.g., \cite{LWGLW}). The paraxial approximation is used for evaluation of the average number of radiated photons. Therefore, the insets with the plots of $dP/dk_0dk_\perp$ versus $n_\perp$ are valid only for small $n_\perp$. On the left panel: The radiation from the one ion. On the right panel: The radiation from the Gaussian collimated isoenergetic beam of ions. The longitudinal size of the beam $\s_3=150$ $\mu$m, the transverse size of the beam $\s_\perp=1$ $\mu$m and the number of particles in the beam $N=10^5$. As long as $k_0\s_3\gg1$, the coherent contribution to radiation from the beam is strongly suppressed. The quantity $\ell$ is the average projection of the total angular momentum per photon. The plots with the photon helicity $s=1$ are not presented because the contribution of such photons to the radiation is negligible. We see that the radiation at given harmonic is a pure sources of twisted photons with $m=-2$, $s=-1$, and $l=m-s=-1$, in accordance with the theory.}}
\label{U_parax_plots}
\end{figure}

It is not difficult to see from Eqs. \eqref{Max_eqns3} that the following estimates take place
\begin{equation}
    a_+(p_3+2ql)=O(k_\perp^{2|l|}),\qquad a_-(p_3-2q+2ql)=O(k_\perp^{2|l|}).
\end{equation}
Hence, keeping only the first order terms in $k_\perp^2$, we obtain for the mode $1$:
\begin{equation}\label{disper_parax1}
    p^{(1)}_3=p^{(1)}_0-\frac{k_\perp^2}{2\bar{k}_0 wv_+} \Big[\Big(1+\frac{\de\e}{4}\Big)\Big(1+\frac{\bar{k}_0w}{q}\Big)+\frac{\de\e^2\bar{k}_0^2}{16q^2}\Big].
\end{equation}
The coefficients of the Fourier series \eqref{Fourier_ser} of this mode are given in \eqref{a1_kp}. As far as the second branch is concerned, we have
\begin{equation}\label{disper_parax2}
    p^{(2)}_3=p^{(2)}_0+\frac{k_\perp^2}{2\bar{k}_0 wv_-} \Big[\Big(1+\frac{\de\e}{4}\Big)\Big(1-\frac{\bar{k}_0w}{q}\Big)+\frac{\de\e^2\bar{k}_0^2}{16q^2}\Big].
\end{equation}
The coefficients of the Fourier series \eqref{Fourier_ser} of the mode $2$ are presented in \eqref{a2_kp}. The general expression for the mode function components is written as
\begin{equation}\label{mode_gen}
    a^{(i)}_\pm(z)=\sum_{l=-\infty}^\infty a^{(i)}_\pm(p^{(i)}_3+2ql) e^{i(p^{(i)}_3+2ql)\bar{\theta}/q},\quad i=\overline{1,2}.
\end{equation}
The third component is calculated by the use of formula \eqref{a_3_comp}. The rest two modes $b^{(i)}_\pm$ describing the reflected waves are deduced from the modes $a_\pm^{(i)}$ by means of the symmetry transform \eqref{symm_transf}.

Notice that the expressions for the obtained mode functions comply with the symmetry property mentioned in \cite{BKL5}, the particular case of which the CLCs are. Having performed the Fourier transform over $\vf$ as in passing to the twisted photon mode functions \eqref{pln_waves_to_twisted}, the resulting new mode functions have the form given in Eqs. (129), (131) of \cite{BKL5}. It is a consequence of the fact that the dependence on $z$ of the mode functions \eqref{mode_gen} is gathered into the combination $\bar{\theta}$. However, as we shall see below, this property is violated by taking into account the reflection of the electromagnetic wave at the boundaries of the CLC plate.

Demanding that the higher order corrections with respect to $k_\perp^2$ are small in comparison with the main contribution, we infer that the perturbation theory with respect to $k_\perp$ is applicable provided that
\begin{equation}\label{applic_conds_parax}
\begin{aligned}
    \frac{k_\perp^2}{\bar{k}_0^2}&\ll1,&\qquad\frac{k_\perp^2}{\bar{k}_0^2}\frac{|q|}{|\bar{k}_0-q|}&\ll1, \qquad\frac{k_\perp^2}{|\bar{k}_0-q||\bar{k}_0-2q|}\frac{\de\e^2\bar{k}_0^2}{256 q^2}\ll1,&\quad&\text{for $\frac{\de\e^2\bar{k}_0^2}{16 q^2}\ll1$};\\
    \frac{k_\perp^2}{\bar{k}_0^2}&\ll1,&\qquad\frac{|\de\e|\bar{k}_0}{16|q|}\frac{k_\perp^2}{\bar{k}_0^2}&\ll1,& \quad&\text{for $\frac{\de\e^2\bar{k}_0^2}{16 q^2}\gg1$};
\end{aligned}
\end{equation}
where it is assumed that $|\de\e|\ll1$. As expected, the developed perturbation theory with respect to $k_\perp^2$ does not work near the Bragg resonances. This is the result of the fact that we expanded the dispersion relation in $k_\perp$. In order to construct the mode functions near the Bragg resonances, the nonperturbative dispersion relation following from \eqref{Max_eqns3} in some $n$-wave approximation is needed. Furthermore, the region of applicability of the perturbation theory with respect to $k_\perp^2$ is narrowed in the case of hard photons due to the second condition on the second line in \eqref{applic_conds_parax}. The approximate dispersion laws for $|\de\e|\ll1$ and $|\de\e|\bar{k}_0/4\ll|q|$ read
\begin{equation}
    p_3^{(1)}=2q+\Big(1+\frac{\de\e}{4}\Big)\bar{k}_0-\frac{k_\perp^2}{2\bar{k}_0}+\cdots,\qquad
    p_3^{(2)}=
    \left\{
      \begin{array}{ll}
        \Big(1+\frac{\de\e}{4}\Big)\bar{k}_0-\frac{k_\perp^2}{2\bar{k}_0}+\cdots, & \hbox{\text{for $\bar{k}_0>q$};} \\
        2q-\Big(1+\frac{\de\e}{4}\Big)\bar{k}_0+\frac{k_\perp^2}{2\bar{k}_0}+\cdots, & \hbox{\text{for $\bar{k}_0\in(0,q]$}.}
      \end{array}
    \right.
\end{equation}
For $\de\e>0$, $\de\e\ll1$, and $\de\e\bar{k}_0/4\gg|q|$, they become
\begin{equation}\label{disper_law_parHE}
    p_3^{(1)}=q+\e_\parallel^{1/2}k_0-\frac{k_\perp^2}{2\e_\parallel^{1/2}k_0}+\cdots,\qquad p_3^{(2)}=q+\bar{k}_0-\frac{k_\perp^2}{2\bar{k}_0}+\cdots.
\end{equation}
When $\de\e<0$, $|\de\e|\ll1$, and $|\de\e|\bar{k}_0/4\gg |q|$, the approximate expressions for the dispersion laws take the same form but with the replacement $p_3^{(1)}\leftrightarrow p_3^{(2)}$.

Now we have to join the solutions of the Maxwell equations in the CLC plate with the free electromagnetic waves in vacuum at $z>0$ and $z<-L$ using the boundary conditions \eqref{bound_conds_chol}. For $z>0$ the mode function of the quantum electromagnetic field, $\bs{\psi}(s,\spk;x)$, is given by
\begin{equation}\label{mode_funcz>0}
    \frac{c}{\sqrt{2k_0V}}\mathbf{f}(s,\spk)e^{-ik_0x^0+i\spk\spx}.
\end{equation}
In the CLC plate, it is written as
\begin{equation}\label{mode_func_cho}
    \frac{c}{\sqrt{2k_0V}}[r_1\bs{\psi}^{(1)}_r+r_2\bs{\psi}^{(2)}_r+l_1\bs{\psi}^{(1)}_l+l_2\bs{\psi}^{(2)}_l]e^{-ik_0 x^0+i\spk_\perp \spx_\perp},
\end{equation}
where the components $\bs{\psi}^{(i)}_r$ and $\bs{\psi}^{(i)}_l$, $i=\overline{1,2}$, have the form \eqref{mode_gen} in the basis \eqref{cylin_basis} with the Fourier coefficients $a^{(i)}(\cdot)$ and $b^{(i)}(\cdot)$, respectively. Recall that one should bear in mind the replacement \eqref{apm_not} in order to obtain the components of the mode functions. For $z<-L$, the mode function is written as
\begin{equation}\label{mode_funcz<mL}
    \frac{c}{\sqrt{2k_0V}}\big\{[d_+\mathbf{f}_{++} +d_-\mathbf{f}_{-+}]e^{ik_3z} +[h_+\mathbf{f}_{+-} +h_-\mathbf{f}_{--}]e^{-ik_3z}\big\}e^{-ik_0x^0+ik_\perp\spx_\perp},
\end{equation}
where
\begin{equation}
    \mathbf{f}_{++}:=\mathbf{f}(1,\spk),\qquad \mathbf{f}_{-+}:=\mathbf{f}(-1,\spk),\qquad \mathbf{f}_{+-}:=\mathbf{f}(1,\spk_\perp,-k_3),\qquad \mathbf{f}_{--}:=\mathbf{f}(-1,\spk_\perp,-k_3).
\end{equation}
The constant $c$ is found from the normalization condition (see below). In fact, the solution \eqref{mode_funcz>0}-\eqref{mode_funcz<mL} is the Jost function $F^+(s,k_3)$ defined in \eqref{Jost_funcs}.

It is useful to write the joining conditions \eqref{bound_conds_chol} as the matrix equation
\begin{equation}\label{joining_eqn}
    \left[
       \begin{array}{cc}
         U & 0 \\
         UT & -U_0T' \\
       \end{array}
     \right]
     \left[
       \begin{array}{c}
         a_{ch} \\
         a_l \\
       \end{array}
     \right]=
     \left[
       \begin{array}{c}
         g \\
         0 \\
       \end{array}
     \right],
\end{equation}
where
\begin{equation}
    a^T_{ch}=(r_1,r_2,l_1,l_2),\qquad a_l^T=(d_+,d_-,h_+,h_-).
\end{equation}
Also
\begin{equation}\label{joining_eqn_3}
\begin{gathered}
    T=\diag(e^{-ip_3^{(1)}L},e^{-ip_3^{(2)}L},e^{ip_3^{(1)}L},e^{ip_3^{(2)}L}),\qquad    T'=\diag(e^{-ik_3L},e^{-ik_3L},e^{ik_3L},e^{ik_3L}),\\
    U_0=\frac{1}{\sqrt{2}}
    \left[
      \begin{array}{cccc}
        n_3-1 & n_3+1 & -n_3-1 & -n_3+1 \\
        n_3+1 & n_3-1 & -n_3+1 & -n_3-1 \\
        k_0(1-n_3) & k_0(1+n_3) & k_0(1+n_3) & k_0(1-n_3) \\
        k_0(1+n_3) & k_0(1-n_3) & k_0(1-n_3) & k_0(1+n_3) \\
      \end{array}
    \right],\\
    U=
    \left[
      \begin{array}{cc}
        U_{11} & U_{12} \\
        U_{21} & U_{22} \\
      \end{array}
    \right],\qquad
    U_{11}=
    \left[
      \begin{array}{cc}
        a_+^{(1)} & a_+^{(2)} \\
        a_-^{(1)} & a_-^{(2)} \\
      \end{array}
    \right],\qquad U_{12}=U_{11}\Big|_{a\rightarrow b},\\
    U_{21}=-iK \partial_zU_{11},\qquad U_{22}=-iK \partial_zU_{12}=-U_{21}\big|_{a\rightarrow b},\\
    g^T=\big[n_3-s,n_3+s,k_0(1-sn_3),k_0(1+sn_3)\big]/\sqrt{2}.
\end{gathered}
\end{equation}
In the expressions for the components of the matrix $U$, all the functions of $z$ are taken at $z=0$. Explicitly,
\begin{equation}
\begin{split}
    U_{11}=&\sum_{l=-\infty}^\infty e^{-2il\vf}
    \left[
      \begin{array}{cc}
        a_{l,+}^{(1)} & a_{l,+}^{(2)} \\
        a_{l,-}^{(1)} & a_{l,-}^{(2)} \\
      \end{array}
    \right]
     \diag(e^{-ip^{(1)}_3\vf/q},e^{-ip^{(2)}_3\vf/q}),\\
    U_{21}=&\sum_{l=-\infty}^\infty e^{-2il\vf}
    \left[
      \begin{array}{cc}
        a_{l,+}^{(1)}+\frac{k_\perp^2}{2\bar{k}_3^2}(a_{l,+}^{(1)}+a_{l,-}^{(1)}) & a_{l,+}^{(2)}+\frac{k_\perp^2}{2\bar{k}_3^2}(a_{l,+}^{(2)}+a_{l,-}^{(2)}) \\
        a_{l,-}^{(1)}+\frac{k_\perp^2}{2\bar{k}_3^2}(a_{l,-}^{(1)}+a_{l,+}^{(1)}) & a_{l,-}^{(2)}+\frac{k_\perp^2}{2\bar{k}_3^2}(a_{l,-}^{(2)}+a_{l,+}^{(2)}) \\
      \end{array}
    \right]\times\\
    &\times \diag((p^{(1)}_3+2ql)e^{-ip^{(1)}_3\vf/q},(p^{(2)}_3+2ql)e^{-ip^{(2)}_3\vf/q}),
\end{split}
\end{equation}
where $a^{(i)}_{l,\pm}\equiv a^{(i)}_\pm(p_3^{(i)}+2ql)$. Then we obtain
\begin{equation}\label{joining_sol}
    a_{ch}=U^{-1}g,\qquad a_l=(T')^{-1}U_0^{-1}UTU^{-1}g,
\end{equation}
for the solution of Eq. \eqref{joining_eqn}.

The solution \eqref{joining_sol} possesses certain symmetries. Namely, it obeys the unitarity relations \eqref{unitar1}. In particular,
\begin{equation}
    1+|h_+|^2+|h_-|^2=|d_+|^2+|d_-|^2.
\end{equation}
Besides, if one multiplies the matrix $U$ by the diagonal matrix $D$, i.e., $U\rightarrow UD$, then $a_l$ does not change and
\begin{equation}
    a_{ch}\rightarrow D^{-1}a_{ch}.
\end{equation}
Such a transform corresponds to multiplication of every mode $\bs{\psi}^{(i)}_r$, $\bs{\psi}^{(i)}_l$ by its own constant. This transform allows one to take easily into account the factors $e^{\mp ip_3^{(i)}\vf/q}$ in the matrix $U$ in solving Eq. \eqref{joining_eqn}.

The final expression for the obtained mode function $\bs{\psi}(s,\spk;x)$ should be normalized. The normalization condition results in (see for details Sec. 5.A of \cite{BKL5})
\begin{equation}\label{norm_const0}
    |c|^2=2(1+a^\dag_l a_l)^{-1}=\big(|d_+|^2+|d_-|^2\big)^{-1}.
\end{equation}
The explicit expressions for $a_{ch}$, $a_l$, and the normalization constant are given in Appendix \ref{ParAx_App}. In the leading order in $k_\perp$, i.e., at $k_\perp=0$, the dependence of the coefficients of the linear combination on $\vf$ becomes
\begin{equation}\label{bar_r_bar_l}
    r_{1,2}=\bar{r}_{1,2} e^{-i(s+1)\vf}e^{ip^{(1,2)}_3\vf/q},\qquad l_{1,2}=\bar{l}_{1,2} e^{-i(s-1)\vf}e^{-ip^{(1,2)}_3\vf/q},
\end{equation}
where $\bar{r}_{1,2}$ and $\bar{l}_{1,2}$ are independent of $\vf$. The normalization constant $c$ does not depend on $\vf$ in this limit.

\subsection{Radiation probability}\label{RadProbSec}

Using the mode functions \eqref{mode_funcz>0}, \eqref{mode_func_cho}, \eqref{mode_funcz<mL}, it is not difficult to find the average number of plane-wave photons created by a particle with the charge $Ze$ moving uniformly and rectilinearly. The trajectory of such a particle is
\begin{equation}\label{trajectory}
    \spx=\bs{\be}t,\qquad x^0=t.
\end{equation}
The average number of photons produced by the charged particle is written as
\begin{equation}\label{dP_plane}
    dP(s,\spk)=Z^2e^2\Big|\int_{-\infty}^\infty dt\dot{\spx}(t)\bs{\psi}(s,\spk;x(t))\Big|^2 \frac{Vd\spk}{(2\pi)^3}.
\end{equation}
The inclusive probability to detect a photon by the detector is also determined by the integral entering into \eqref{dP_plane} (see for details \cite{BKL2,BKL5,ippccb}).

\begin{figure}[tp]
\centering
\includegraphics*[width=0.49\linewidth]{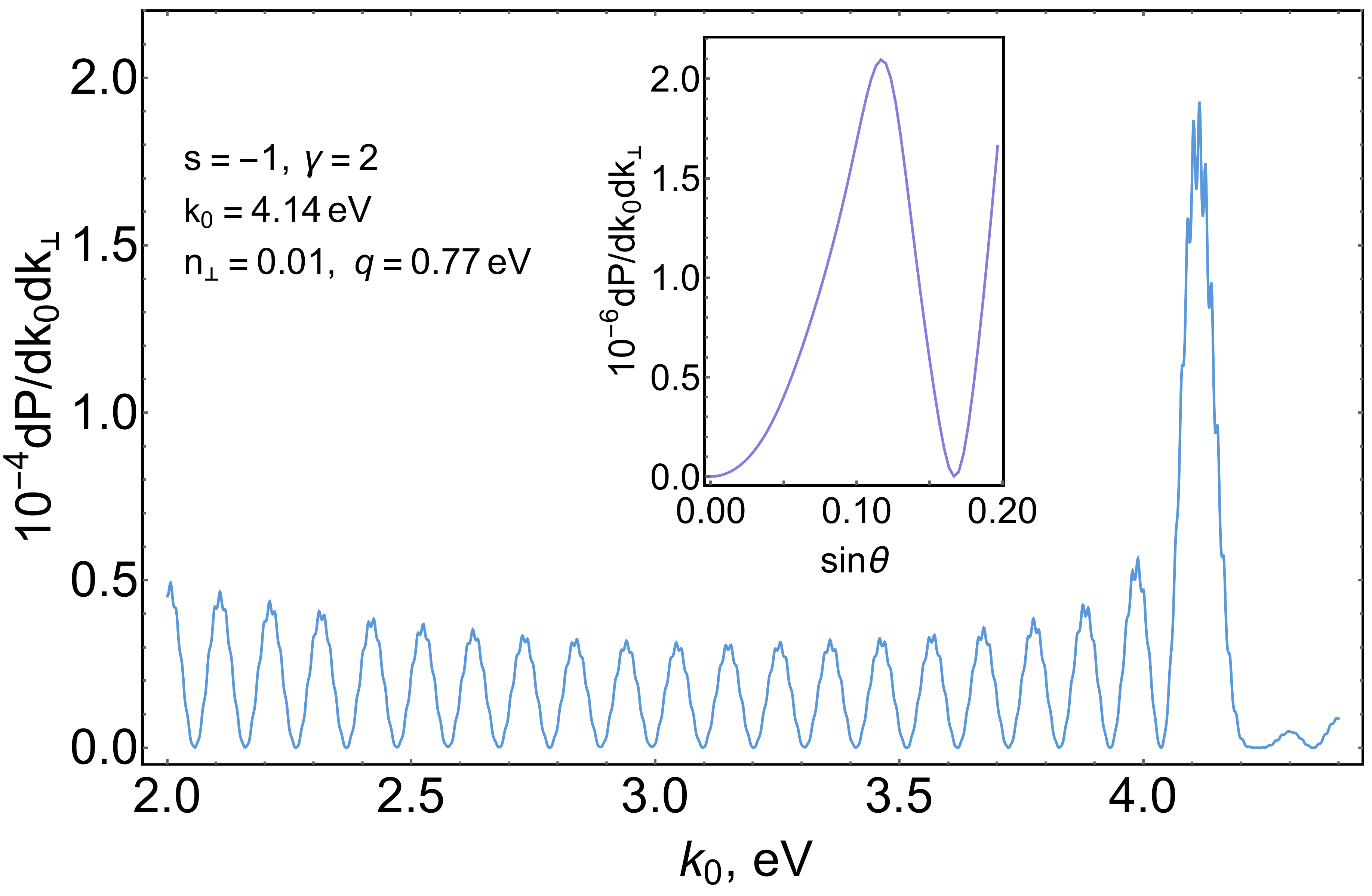}\,
\includegraphics*[width=0.49\linewidth]{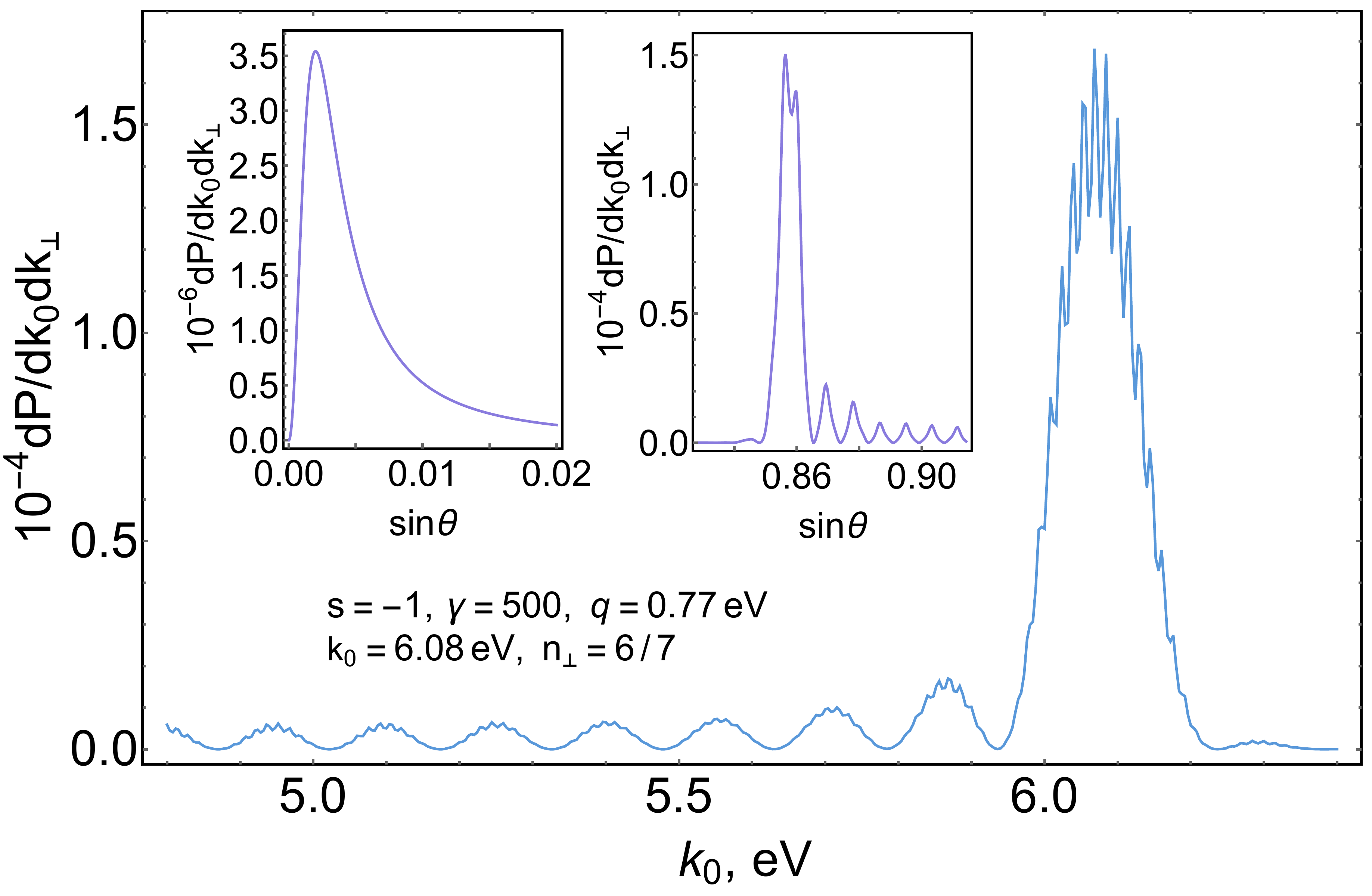}
\caption{{\footnotesize The average number of plane-wave photons produced in transition radiation from the CLC plate. The parameters of the CLC plate and the particle beam are the same as in Fig. \ref{U_parax_plots}. As long as the coherent contribution to radiation is suppressed for such a beam of charged particles, the average number of plane-wave photons produced by the collimated isoenergetic beam differs from the same quantity for one particle by the factor $N$ only. It turns out that the radiation of photons with $s=1$ is strongly suppressed near the main peak. Therefore, the plots for $s=-1$ are only depicted. On the left panel: Transition radiation from the ion ${}^{238}$U${}^{92+}$ traversing normally the CLC plate with the Lorentz factor $\ga=2$. The paraxial approximation is used. The series of peaks on the left from the main peak on the plot with respect to the photon energy is the contribution of the edge radiation from the interfaces of the CLC plate. On the right panel: Transition radiation from the electron traversing normally the CLC plate with the Lorentz factor $\ga=500$, $E\approx 256$ MeV. The small anisotropy approximation is employed. The large peak at small $n_\perp$ on the left inset corresponds to the edge radiation.}}
\label{parax_anis_plw_plots}
\end{figure}

The expression under the modulus sign, which is proportional to the one-particle amplitude, is the sum of several terms: the edge radiation from the part of the trajectory $z>0$,
\begin{equation}\label{edge_rad>0}
    \frac{c}{\sqrt{2k_0V}}\frac{i\bs{\beta}\mathbf{f}(s,\spk)}{k_0(1-\mathbf{n}\bs{\beta})};
\end{equation}
the edge radiation from the part of the trajectory $z<-L$,
\begin{equation}\label{edge_rad<mL}
    \frac{c}{\sqrt{2k_0V}}\Big[\frac{\bs{\be}(\mathbf{f}_{++}d_+ +\mathbf{f}_{+-}d_-)}{k_0(1-\mathbf{n}\bs{\be})}ie^{ik_0(1-\mathbf{n}\bs{\be})L/\be_3} +\frac{\bs{\be}(\mathbf{f}_{-+}h_+ +\mathbf{f}_{--}h_-)}{k_0(1-\mathbf{n}_\perp\bs{\be}_\perp+n_3\be_3)}ie^{ik_0(1-\mathbf{n}_\perp\bs{\be}_\perp+n_3\be_3)L/\be_3} \Big];
\end{equation}
and the contribution of the transition radiation from the periodic permittivity in the cholesteric plate that is given by the integral over $t$ from $-L/\be_3$ to $0$ of the expression
\begin{multline}\label{scal_prod_bp}
    \bs{\beta}\bs{\psi}(s,\spk;x(t))=\sum_{n=-\infty}^\infty\sum_{i=1}^2 \bigg[ \frac{r_i}{2}\Big\{\bar{\be}_+a^{(i)}_{n,-} +\bar{\be}_-a^{(i)}_{n,+} -\frac{k_\perp}{\bar{k}_3^2}\be_3(p_3^{(i)}+2qn)(a^{(i)}_{n,+}+a^{(i)}_{n,-})\Big\}e^{i(p_3^{(i)}+2qn) (\be_3t-\vf/q)}+\\
    +\frac{l_i}{2}\Big\{\bar{\be}_+a^{(i)}_{n,+} +\bar{\be}_-a^{(i)}_{n,-} +\frac{k_\perp}{\bar{k}_3^2}\be_3(p_3^{(i)}+2qn)
    (a^{(i)}_{n,+}+a^{(i)}_{n,-})\Big\}e^{-i(p_3^{(i)}+2qn) (\be_3t-\vf/q)} \bigg] \frac{c  e^{-ik_0t(1-\mathbf{n}_\perp\bs{\be}_\perp)}}{\sqrt{2k_0V}},
\end{multline}
where $\bar{\be}_\pm:=\be_\pm e^{\mp i\vf}$ and $\be_\perp=|\be_+|$.

Hereinafter we will be interested in the case when $N_u\gg1$. Notice that the number of periods, $N_u$, in the usual CLCs cannot be too large and, in fact, is bounded from above as $N_u\lesssim40$ \cite{DenKizSukh}. Otherwise, the CLC helices are distorted in the volume of the cholesteric. For $N_u$ sufficiently large, the main contribution to the radiation amplitude comes from the last integral. Neglecting the edge radiation, we arrive at the following expression for the one-particle amplitude of photon radiation:
\begin{equation}\label{amplitude_parax}
\begin{split}
    \frac{Zec}{2\sqrt{2k_0V}}\sum_{n=-\infty}^\infty\sum_{i=1}^2 &\bigg[\vf(x_n^{(i)}) \bar{r}_i\Big\{\bar{\be}_+a^{(i)}_{n,-} +\bar{\be}_-a^{(i)}_{n,+}
    -\frac{k_\perp}{\bar{k}_3^2}\be_3(p_3^{(i)}+2qn) (a^{(i)}_{n,+} +a^{(i)}_{n,-})\Big\}e^{-i(2n+1+s)\vf}+\\
    &+\vf(\tilde{x}_n^{(i)}) \bar{l}_i\Big\{\bar{\be}_+a^{(i)}_{n,+} +\bar{\be}_-a^{(i)}_{n,-} +\frac{k_\perp}{\bar{k}_3^2}\be_3(p_3^{(i)}+2qn)
    (a^{(i)}_{n,+} +a^{(i)}_{n,-})\Big\} e^{i(2n+1-s)\vf} \bigg],
\end{split}
\end{equation}
where
\begin{equation}\label{xn_tilde_xn}
\begin{gathered}
    x^{(i)}_n=k_0(1-\mathbf{n}_\perp\bs{\be}_\perp)-\be_3(p_3^{(i)}+2qn),\qquad \tilde{x}^{(i)}_n=k_0(1-\mathbf{n}_\perp\bs{\be}_\perp)+\be_3(p_3^{(i)}+2qn),\\
    \vf(x):=2\pi e^{iTN_ux/2}\de_{N_u}(x),\qquad \de_{N_u}(x):=\frac{\sin(TN_ux/2)}{\pi x},\qquad T:=\pi/(|q|\be_3).
\end{gathered}
\end{equation}
For $N_u\gg1$, the modulus of the function $\vf(x)$ possesses a sharp maximum at $x=0$. Therefore, the main contribution to the radiation amplitude comes from the harmonics $x^{(i)}_n=0$ and $\tilde{x}^{(i)}_n=0$. In the paraxial approximation we consider, the main contribution to the amplitude is provided by the harmonics with the numbers $n=\overline{-2,1}$ for the modes at the coefficients $r_i$ and by the harmonics with the numbers $n=\{-2,-1\}$ for the modes at the coefficients $l_i$. The terms with $n=\{-1,0\}$ dominate.

Usually the CLC is not placed in the vacuum but is confined into some cuvette made of a transparent medium. The cuvette exerts influence on the solutions of the Maxwell equations. However, if this cuvette is transparent in the given spectral range, is made of isotropic homogenous material, and its faces are parallel to each other, then it does not change substantially the contribution to radiation produced by charged particles in the CLC, which is proportional to $N_u^2$.

Thus, in the leading order, the average number of radiated plane-wave photons reads
\begin{equation}\label{aver_num_phot}
\begin{split}
    dP(s,\spk)=&\,|Zec|^2\sum_{n=-\infty}^\infty\sum_{i=1}^2 \bigg[\de_{N_u}^2(x_n^{(i)}) |\bar{r}_i|^2\Big|\bar{\be}_+a^{(i)}_{n,-} +\bar{\be}_-a^{(i)}_{n,+}
    -\frac{k_\perp}{\bar{k}_3^2}\be_3(p_3^{(i)}+2qn) (a^{(i)}_{n,+} +a^{(i)}_{n,-})\Big|^2+\\
    &+\de_{N_u}^2(\tilde{x}_n^{(i)}) |\bar{l}_i|^2\Big|\bar{\be}_+a^{(i)}_{n,+} +\bar{\be}_-a^{(i)}_{n,-} +\frac{k_\perp}{\bar{k}_3^2}\be_3(p_3^{(i)}+2qn)
    (a^{(i)}_{n,+} +a^{(i)}_{n,-})\Big|^2 \bigg]\frac{d\spk}{16\pi k_0}.
\end{split}
\end{equation}
The conditions $x^{(i)}_n=0$, $\tilde{x}^{(i)}_n=0$ for the maximum of radiation intensity can be simplified in the case when $|\de\e|\bar{k}_0/4\gg |q|$. Employing the approximate expressions for the dispersion laws \eqref{disper_law_parHE} and introducing the notation
\begin{equation}
    \bar{n}_{3\perp}:=\sqrt{\e_{\perp}-n_\perp^2}\approx \e_{\perp}^{1/2}-n_\perp^2/(2\e_{\perp}^{1/2}),\qquad \bar{n}_{3\parallel }:=\sqrt{\e_{\parallel}-n_\perp^2}\approx \e_{\parallel}^{1/2}-n_\perp^2/(2\e_{\parallel}^{1/2}),
\end{equation}
the radiation spectrum becomes
\begin{equation}\label{spectrum_parax}
    k_0=\frac{\mp\be_3q(2n+1)}{1-\mathbf{n}_\perp\bs{\be}_\perp \pm \bar{n}_{3\perp}\be_3},\qquad k_0=\frac{\mp\be_3q(2n+1)}{1-\mathbf{n}_\perp\bs{\be}_\perp \pm \bar{n}_{3\parallel}\be_3}\quad k_0>0,
\end{equation}
where the upper sign is taken for the solutions to the equation $\tilde{x}^{(i)}_n=0$ and the lower sign is for $x^{(i)}_n=0$. The number of the Fourier harmonic $n$ and the vector $\mathbf{n}$ are selected in such a way that $k_0$ is positive. The plots of the average number of photons produced in transition radiation from uranium nuclei and electrons are given in Fig. \ref{parax_anis_plw_plots}.

Notice that if one defines the Cherenkov cone by analogy with the case of a homogeneous dispersive medium,
\begin{equation}
    1-\mathbf{n}_\perp\bs{\be}_\perp - \bar{n}_{3\perp}\be_3=0,\qquad 1-\mathbf{n}_\perp\bs{\be}_\perp - \bar{n}_{3\parallel}\be_3=0,
\end{equation}
then the radiation does not form at $\mathbf{n}$ satisfying one of these equations since the numerator of \eqref{spectrum_parax} does not vanish for any harmonic number $n$. This property is a consequence of periodicity of the medium permittivity \cite{Ginzburg,BazylZhev,BelDmitOrl}. In fact, at a given photon energy there is an infinite number of Cherenkov cones. The principal Cherenkov cone can be uniquely defined in the case when the perturbation theory with respect to $\de\e$ is applicable (see \eqref{VC_cone}).

In the ultrarelativistic limit, $\ga\gg1$, assuming $\be_\perp\ll1$ and $|\de\e|\bar{k}_0/4\gg |q|$, it follows from \eqref{spectrum_parax} that the radiation spectrum is written as
\begin{equation}\label{spectrum_parax1}
\begin{split}
    k_0&=\frac{2\ga^2q(2n+1)\e_{\perp}^{-1/2}}{1+\ga^2\big(\bs{\beta}_\perp-\e_{\perp}^{-1/2}\mathbf{n}_\perp\big)^2 +2\ga^2\big(\e_{\perp}^{-1/2}-1\big)} \approx\frac{2\ga^2q(2n+1)}{1+\ga^2(\bs{\beta}_\perp-\mathbf{n}_\perp)^2 -\ga^2\chi_{\perp}},\\
    k_0&=\frac{2\ga^2q(2n+1)\e_{\parallel}^{-1/2}}{1+\ga^2\big(\bs{\beta}_\perp-\e_{\parallel}^{-1/2}\mathbf{n}_\perp\big)^2 +2\ga^2\big(\e_{\parallel}^{-1/2}-1\big)} \approx\frac{2\ga^2q(2n+1)}{1+\ga^2(\bs{\beta}_\perp-\mathbf{n}_\perp)^2 -\ga^2\chi_{\parallel}},\\
    k_0&=-\frac{q(2n+1)}{1+\e_{\perp}^{1/2}},\qquad k_0=-\frac{q(2n+1)}{1+\e_{\parallel}^{1/2}},
\end{split}
\end{equation}
where it is supposed in the approximate equality that the modulus of $\chi_{\perp,\parallel}:=\e_{\perp,\parallel}-1$ is much less than unity. The expressions on the third line in \eqref{spectrum_parax1} correspond to the equation $\tilde{x}^{(i)}_n=0$ and can be satisfied only for negative $n$.

Now it is not difficult to obtain the average number of radiated twisted photons in the case $\bs{\be}_\perp=0$. Expanding the scalar product \eqref{amplitude_parax} in terms of the twisted photons with the help of \eqref{pln_waves_to_twisted}, we derive the one-particle amplitude of radiation of a twisted photon
\begin{equation}\label{ampl_tw_parax}
    \frac{Zec\be_3}{4\sqrt{RL_z}}\frac{n_\perp^{3/2}}{i^m\e_\perp k_0}\sum_{n=-\infty}^\infty\sum_{i=1}^2 \big[\vf(y_n^{(i)}) \bar{r}_i\de_{m,s+2n+1}-\vf(\tilde{y}_n^{(i)}) \bar{l}_i \de_{m,s-2n-1}\big]
     (p_3^{(i)}+2qn)(a^{(i)}_{n,+}+a^{(i)}_{n,-}),
\end{equation}
in the leading order in $k_\perp^2$, where
\begin{equation}\label{yn_tilde_yn}
    y^{(i)}_n=k_0-\be_3(p_3^{(i)}+2qn),\qquad \tilde{y}^{(i)}_n=k_0+\be_3(p_3^{(i)}+2qn).
\end{equation}
The energy spectrum of radiated twisted photons is specified by the equations $y^{(i)}_n=0$, $\tilde{y}^{(i)}_n=0$, which are obtained from \eqref{xn_tilde_xn} at $\be_\perp=0$. Rising the modulus of \eqref{ampl_tw_parax} to the second power and neglecting the small contributions when $N_u\gg1$, we find the average number of radiated twisted photons
\begin{equation}\label{dP_tw_par}
\begin{split}
    dP(s,m,k_\perp,k_3)=\,&|Zec\be_3|^2 \sum_{n=-\infty}^\infty\sum_{i=1}^2 \big[\de_{N_u}^2(y_n^{(i)}) |\bar{r}_i|^2\de_{m,s+2n+1}+\de_{N_u}^2(\tilde{y}_n^{(i)}) |\bar{l}_i|^2 \de_{m,s-2n-1}\big]\times\\
    &\times(p_3^{(i)}+2qn)^2(a^{(i)}_{n,+} +a^{(i)}_{n,-})^2 n_\perp^3 \frac{dk_3 dk_\perp}{8\e_\perp^2 k_0^2}.
\end{split}
\end{equation}
As is seen, for the photon energies satisfying $y^{(i)}_n=0$, the following selection rule for the projection of the orbital angular momentum is fulfilled:
\begin{equation}\label{rm_spectr}
    l:=m-s=2n+1.
\end{equation}
Whereas for $\tilde{y}^{(i)}_n=0$, the selection rule,
\begin{equation}\label{lm_spectr}
    l=-2n-1,
\end{equation}
holds. The plots of the average number of twisted photons created in transition radiation from uranium nuclei in the paraxial regime are presented in Fig. \ref{U_parax_plots}.

Thus we see that the system under study is a pure source of twisted photons in the paraxial regime. Notice that the radiation possesses a nonzero projection of the orbital angular momentum for any $n$. Besides, in both cases \eqref{rm_spectr}, \eqref{lm_spectr}, the projection of the total angular momentum $m$ is an even number. For the approximation considered, the terms of the Fourier series with $n=\{-1,0\}$ dominate. Therefore, for the harmonics, where \eqref{rm_spectr} is fulfilled, we have $l=\{-1,1\}$ in general. As for the harmonics \eqref{lm_spectr}, the projection of the orbital angular momentum of radiated twisted photons is $l=1$. Recall that we take $q>0$. It is clear that the signs of $m$, $s$, and $l$ of the radiated photons are flipped when the sign of $q$ is changed and all the other parameters are left intact.

\section{Perturbation theory with respect to anisotropy}\label{AnisotrAppr}

Now we consider the perturbative solution of Eqs. \eqref{Max_eqns3} regarding $\de\e$ as a small parameter. It is not difficult to see from Eqs. \eqref{Max_eqns3} that for such a perturbation theory
\begin{equation}
    a_\pm(p_3+2ql)=O(\de\e^{2|l|}).
\end{equation}
The dispersion law for the mode $1$ has the form
\begin{equation}
    p^{(1)}_3=\bar{k}_3\Big(1+\frac{\de\e}{4} -\frac{\de\e^2}{64}\frac{4\bar{k}_3^2+k_\perp^2-2q^2}{\bar{k}_3^2-q^2}\Big).
\end{equation}
The coefficients of the Fourier series \eqref{mode_gen} are presented in \eqref{a1_de} up to the terms of the order $\de\e$. Analogously, we obtain for the mode $2$:
\begin{equation}
    p^{(2)}_3=\bar{k}_3\Big(1+\frac{\de\e}{4}\frac{\bar{k}_0^2}{\bar{k}_3^2} -\frac{\de\e^2}{64}\bar{k}_0^2\frac{4 \bar{k}_3^4+\bar{k}_3^2 (3 k_\perp^2-2q^2)-2k_\perp^2}{\bar{k}_3^4(\bar{k}_3^2-q^2)}\Big).
\end{equation}
The expressions for the coefficients of the Fourier series \eqref{mode_gen} for this mode are given in \eqref{a2_de}.

\begin{figure}[tp]
\centering
\includegraphics*[width=0.49\linewidth]{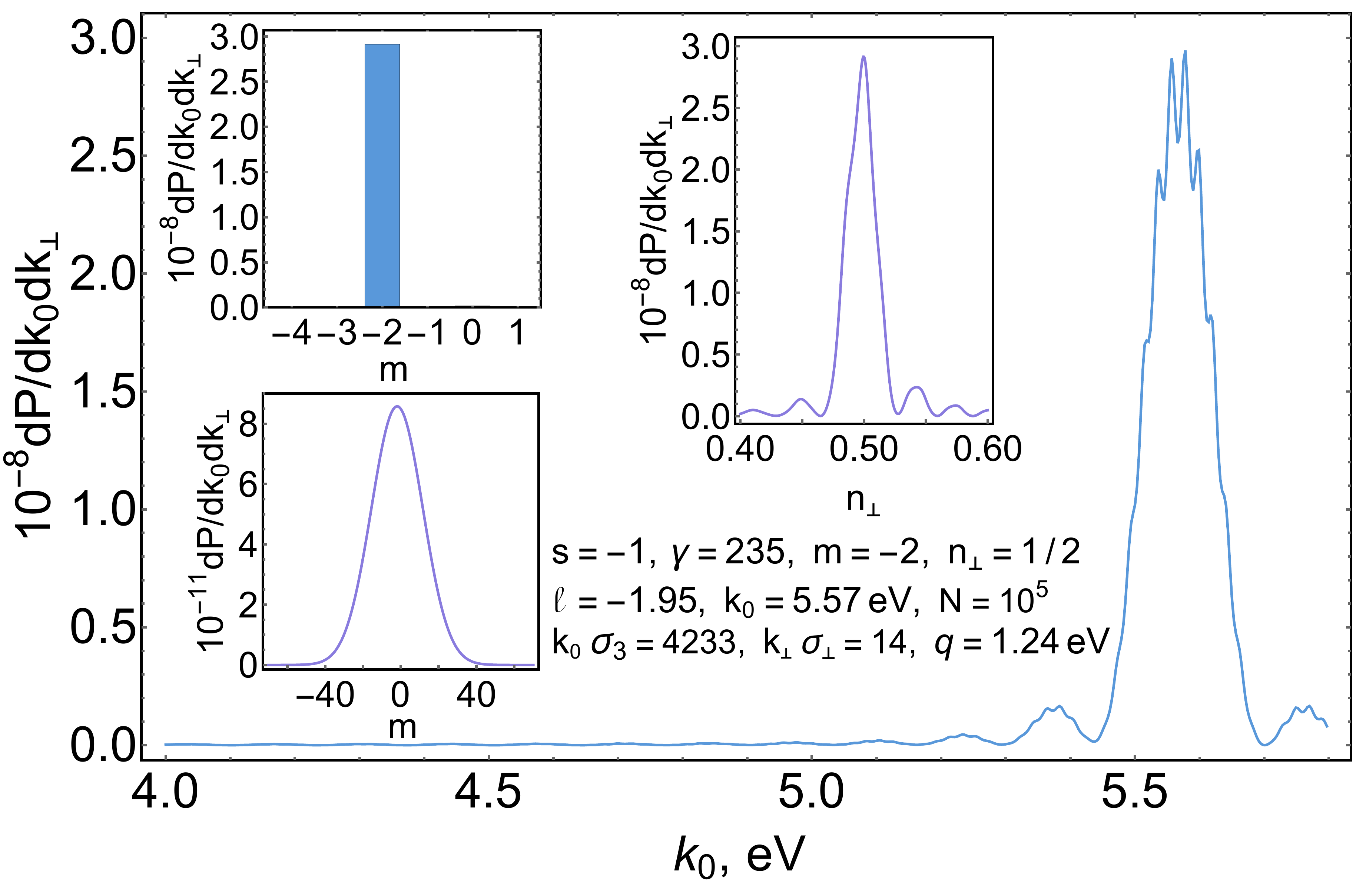}\,
\includegraphics*[width=0.49\linewidth]{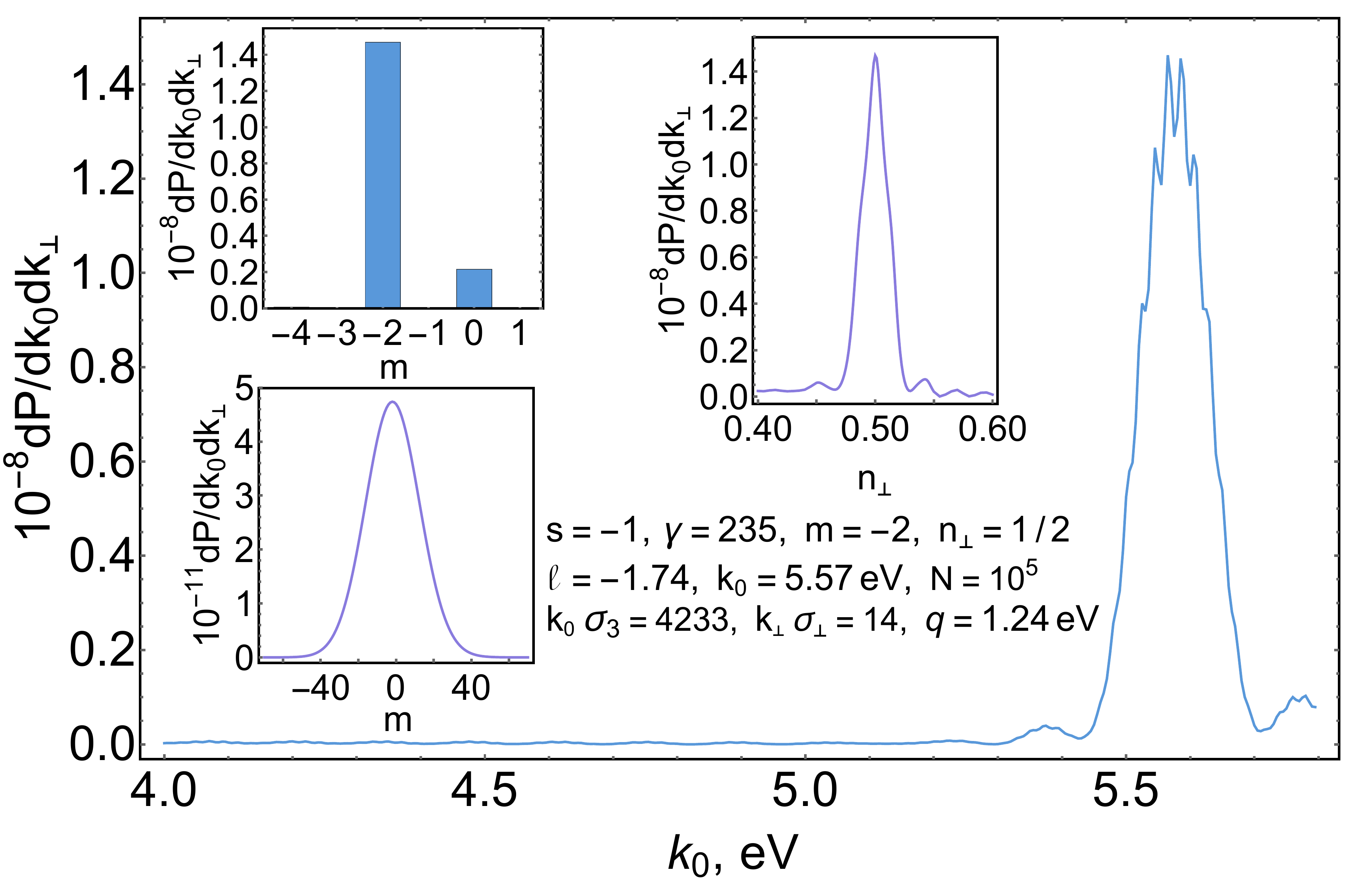}
\caption{{\footnotesize The comparison of the paraxial and the small anisotropy approximations in the parameter space where the both approximation are applicable. The average number of twisted photons produced in transition radiation from the CLC plate by the Gaussian beam of electrons with the Lorentz factor $\ga=235$, $E\approx 120$ MeV, is considered. The parameters of the beam and of the permittivity tensor $\e_{\perp,\parallel}$ are the same as in Fig. \ref{U_parax_plots}. The CLC plate has the width $L=20$ $\mu$m, the number of periods $N_u=40$ corresponding to $q\approx1.24$ eV, and the CLC helix pitch $1$ $\mu$m. The radiation of photons with $s=1$ is suppressed. The distributions over the projections of the total angular momentum $m$ of twisted photons radiated by the one electron are presented on the top left insets. The same distributions for twisted photons radiated by the beam of electrons are given on the bottom left insets. Since $k_\perp\s_\perp\gg1$, the distribution over $m$ is wide in the case of radiation from the beam. Nevertheless, the average projection of the total angular momentum per photon, $\ell$, is the same as for the one-particle distribution \cite{BKb,BKLb}. On the left panel: The paraxial approximation is used. On the right panel: The small anisotropy approximation is employed. It is seen that the both approximations are in fairly good agreement to each other despite the fact that $n_\perp=1/2$ is not very small.}}
\label{parax_anis_tw_plots}
\end{figure}

\begin{figure}[tp]
\centering
\includegraphics*[width=0.49\linewidth]{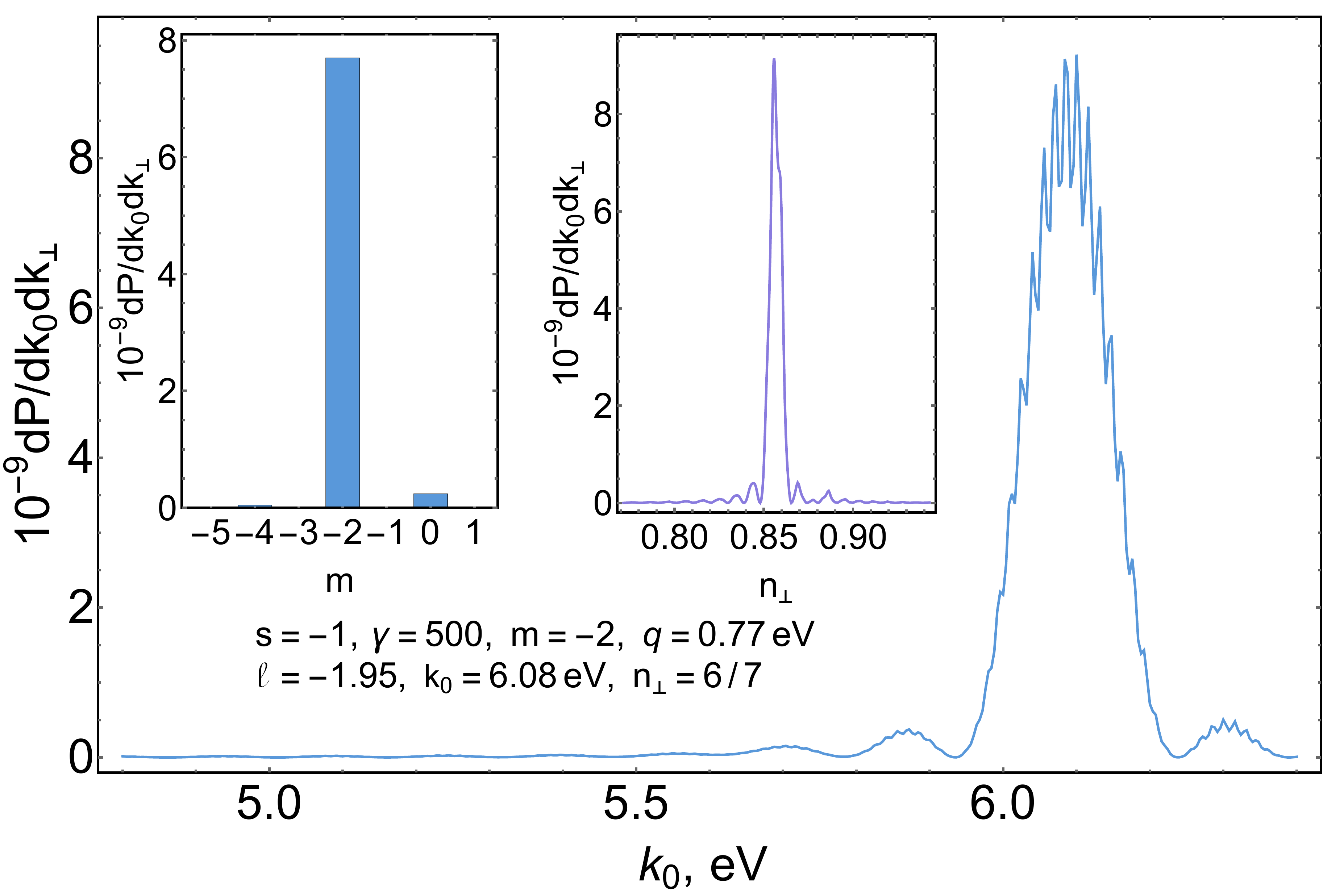}\,
\includegraphics*[width=0.49\linewidth]{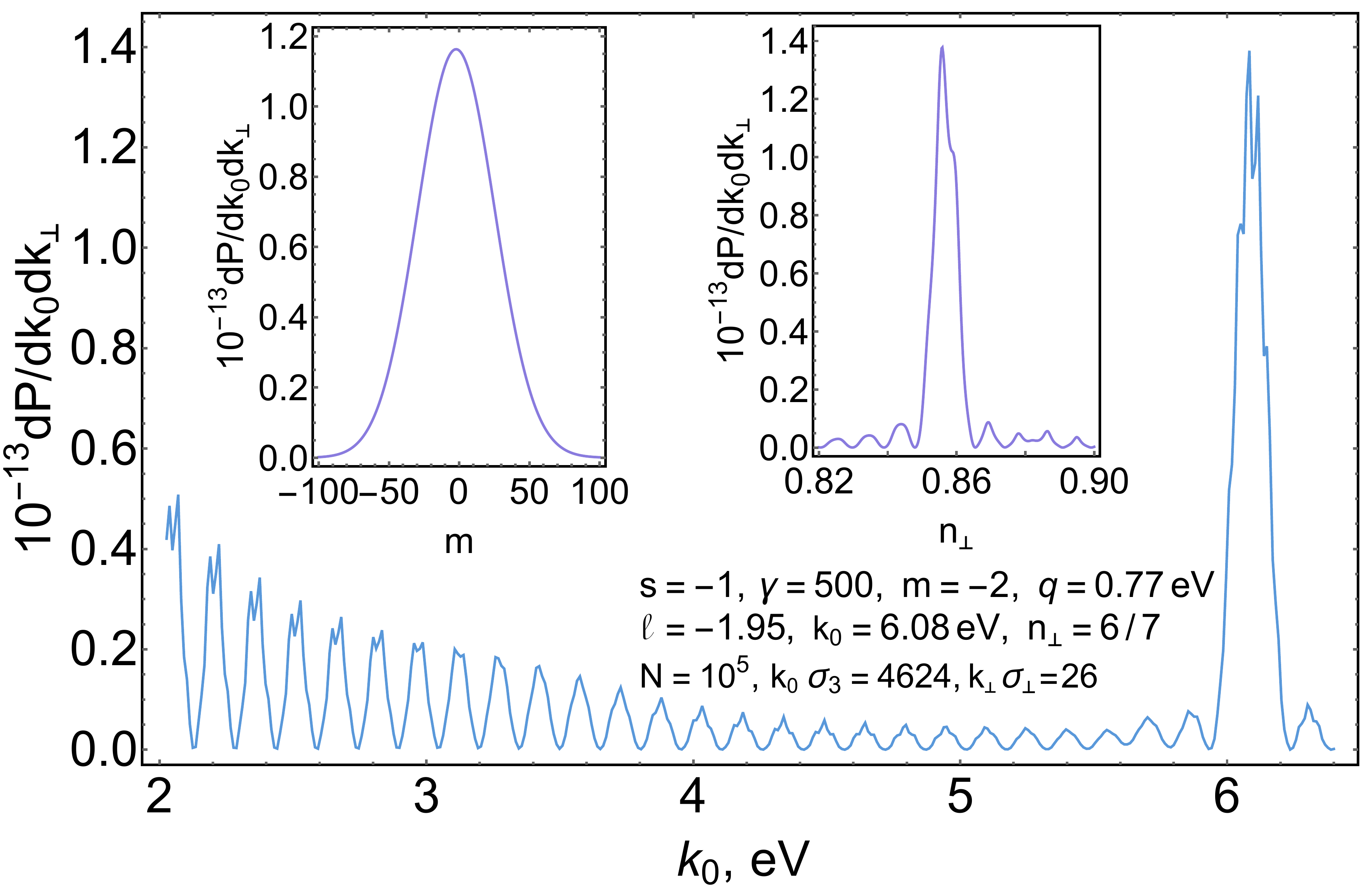}
\caption{{\footnotesize The average number of twisted photons produced in transition radiation from the electrons with the Lorentz factor $\ga=500$, $E\approx 256$ MeV, traversing normally the CLC plate. The small anisotropy approximation is used. The parameters of the CLC plate and the particle beam are the same as in Fig. \ref{U_parax_plots}. The radiation of photons with $s=1$ is suppressed and is not depicted. On the left panel: The radiation from the one electron. The most of twisted photons radiated at the harmonic have $m=-2$ in accordance with the theory. On the right panel: The radiation from the beam of electrons. The series of peaks on the left from the main peak on the plot with respect to the photon energy is the contribution of the edge radiation from the interfaces of the CLC plate. The distribution over $m$ is wide because $k_\perp\s_\perp\gg1$, but the average projection of the total angular momentum per photon, $\ell$, is the same as for the one-particle radiation \cite{BKb,BKLb}.}}
\label{anis_tw_plots}
\end{figure}

It follows from Eqs. \eqref{a1_de} and \eqref{a2_de} that the perturbation theory with respect to $\de\e$ is applicable provided that $|\de\e|\ll1$ and
\begin{equation}\label{appl_cond_de}
    \Big|\frac{\de\e}{4}\frac{\e_\perp}{\e_\perp-n_\perp^2}\Big|\ll1,\qquad \Big|\frac{\de\e\bar{k}^5_3}{4qk_\perp^2 (\bar{k}_3^2-q^2)}\Big|\ll1.
\end{equation}
In particular, the developed perturbation theory in $\de\e$ is not applicable near the Bragg resonances, in the paraxial regime $n_\perp\rightarrow0$, and for the high energies of photons $\de\e k_0\gg |q|$. When $\bar{k}_3^2\gg q^2$, the last condition in \eqref{appl_cond_de} implies
\begin{equation}\label{appl_cond_he_de}
    \Big|\frac{\de\e\bar{k}^3_3}{4qk_\perp^2}\Big|\ll1.
\end{equation}
As is shown in Sec. \ref{RadProbSec}, the relativistic charged particle moving along a straight line in a CLC plate is a pure source of hard twisted photons with definite projection of the orbital angular momentum when $k_0\gg |q|$ and $n_\perp\ll1$. The fulfilment of these conditions tends to violate the applicability condition \eqref{appl_cond_he_de}, and so, as a rule, the main part of radiated hard twisted photons with definite projection of the orbital angular momentum is not described by perturbation theory with respect to $\de\e$. However, for the photon energies much higher than the plasma frequency $\omega_p$, the relative anisotropy $\de\e$ tends to zero as $1/k_0^2$, and the condition \eqref{appl_cond_he_de} becomes less and less stringent. Furthermore, in the optical spectral range, the photons are radiated by a charged particle at sufficiently large angles to the $z$ axis (see Fig.  \ref{parax_anis_plw_plots}, \ref{parax_anis_tw_plots}, \ref{anis_tw_plots}) where the perturbation theory with respect to $\de\e$ can be applicable.

Now we have to join the mode functions of the quantum electromagnetic field in the CLC plate with the corresponding mode functions out of it. The resulting equations for the linear combination coefficients take the form \eqref{joining_eqn}-\eqref{joining_eqn_3}. Their solution is given by formula \eqref{joining_sol}. The explicit expressions for the coefficients $a_{ch}$, $a_l$, and the normalization coefficient $c$ are presented in Appendix \ref{de_App} up to the terms of the order $\de\e$. The normalization coefficient \eqref{c_0} does not depend on $\vf$ in this approximation and the coefficients $\vs_{i}$ \eqref{vsi} entering into $a_{ch}$ \eqref{ach_al_0} are canceled by the corresponding factors in the Fourier series \eqref{mode_gen}.

The one-particle amplitude of photon radiation from a charged particle moving along a straight line \eqref{trajectory} has the form \eqref{amplitude_parax} with the obvious replacements,
\begin{equation}\label{amplitude_de}
\begin{split}
    \frac{Zec}{2\sqrt{2k_0V}}\sum_{n=-\infty}^\infty\sum_{i=1}^2 &\bigg[\vf(x_n^{(i)}) \bar{r}_i\Big\{\bar{\be}_+a^{(i)}_{n,-} +\bar{\be}_-a^{(i)}_{n,+}
    -\frac{k_\perp}{\bar{k}_3^2}\be_3(p_3^{(i)}+2qn)(a^{(i)}_{n,+} +a^{(i)}_{n,-})]\Big\}e^{-2in\vf}+\\
    &+\vf(\tilde{x}_n^{(i)}) \bar{l}_i\Big\{\bar{\be}_+a^{(i)}_{n,+} +\bar{\be}_-a^{(i)}_{n,-} +\frac{k_\perp}{\bar{k}_3^2}\be_3(p_3^{(i)}+2qn)
     (a^{(i)}_{n,+} +a^{(i)}_{n,-})\Big\}e^{2in\vf} \bigg],
\end{split}
\end{equation}
where
\begin{equation}
    \bar{r}_i:=r_i/\vs_{i},\qquad \bar{l}_i:=l_i\vs_{i}.
\end{equation}
For $N_u\gg1$, the radiation is concentrated at the harmonics determined by the equations $x^{(i)}_n=0$, $\tilde{x}^{(i)}_n=0$, where $x^{(i)}_n$ and $\tilde{x}^{(i)}_n$ are defined in \eqref{xn_tilde_xn}. These equations give rise to
\begin{equation}\label{spectrum_de}
    k_0=\frac{\mp2\be_3qn}{1-\mathbf{n}_\perp\bs{\be}_\perp \pm \bar{n}^{(i)}_{3}\be_3},\quad k_0>0,
\end{equation}
where $\bar{n}_3^{(i)}:=p_3^{(i)}/k_0$. The apparent difference of this radiation spectrum from the radiation spectrum \eqref{spectrum_parax} deduced in Sec. \ref{ParaxAppr} stems from the different definition of $\bar{n}_3^{(i)}$. The condition,
\begin{equation}\label{VC_cone}
    1-\mathbf{n}_\perp\bs{\be}_\perp - \bar{n}^{(i)}_{3}\be_3=0,
\end{equation}
corresponds to VC radiation and is realized at $n=0$. If this equation is satisfied for some $\mathbf{n}_\perp$, then this contribution dominates in the radiation. In this case, there are also small contributions from the harmonics with the numbers $n=\overline{-1,1}$ for the modes at the coefficients $r_i$ and with the number $n=-1$ for the modes at the coefficients $l_i$. If the condition \eqref{VC_cone} is not fulfilled for any $n_\perp$, then the contributions with nonpositive $n$ are not realized for the modes at the coefficients $r_i$. The expressions given in \eqref{ach_al_0}, \eqref{c_0} are sufficient to find the leading in $\de\e$ contribution to the amplitude \eqref{amplitude_de}. The average number of radiated plane-wave photons has the form \eqref{aver_num_phot}.

For $\be_\perp=0$, the amplitude of radiation of a one twisted photon is written as
\begin{equation}\label{ampl_tw_de}
    \frac{Zec\be_3}{4\sqrt{RL_z}}\frac{n_\perp^{3/2}}{i^m k_0\bar{n}_3^2}\sum_{n=-\infty}^\infty\sum_{i=1}^2 \big(\vf(y_n^{(i)}) \bar{r}_i\de_{m,2n}-\vf(\tilde{y}_n^{(i)}) \bar{l}_i \de_{m,-2n}\big)
     (p_3^{(i)}+2qn) (a^{(i)}_{n,+} +a^{(i)}_{n,-}),
\end{equation}
where $y_n$ and $\tilde{y}_n$ are defined in \eqref{yn_tilde_yn}. The spectrum of radiation of twisted photons with respect to the energy is determined by the equations $y^{(i)}_n=0$, $\tilde{y}^{(i)}_n=0$ that follow from \eqref{spectrum_de} at $\be_\perp=0$. Squaring the modulus of \eqref{ampl_tw_de} and neglecting the small contributions at $N_u\gg1$, we come to the average number of radiated twisted photons
\begin{equation}\label{dP_tw_anis}
\begin{split}
    dP(s,m,k_\perp,k_3)=\,&|Zec\be_3|^2 \sum_{n=-\infty}^\infty\sum_{i=1}^2 \big(\de_{N_u}^2(y_n^{(i)}) |\bar{r}_i|^2\de_{m,2n}+\de_{N_u}^2(\tilde{y}_n^{(i)}) |\bar{l}_i|^2 \de_{m,-2n}\big)\times\\
    &\times (p_3^{(i)}+2qn)^2 (a^{(i)}_{n,+} +a^{(i)}_{n,-} )^2 n_\perp^3\frac{dk_3 dk_\perp}{8 k_0^2\bar{n}_3^4}.
\end{split}
\end{equation}
For the photon energies $y^{(i)}_n=0$, the selection rule for the projection of the total angular momentum,
\begin{equation}\label{rm_spectr_m}
    m=2n,
\end{equation}
holds. Whereas for $\tilde{y}^{(i)}_n=0$, the selection rule,
\begin{equation}\label{lm_spectr_m}
    m=-2n,
\end{equation}
is fulfilled. In the latter case, $n<0$ because otherwise the condition $\tilde{y}_n=0$ cannot be satisfied. The plots of the average number of planewave and twisted photons produced in transition radiation are given in Figs. \ref{parax_anis_plw_plots}, \ref{parax_anis_tw_plots}, \ref{anis_tw_plots} in the small anisotropy approximation.

As we see, the system under study is a pure source of twisted photons in the range of applicability of perturbation theory with respect to $\de\e$. In the case when VC radiation is not formed, the main contribution to radiation comes from the twisted photons with $m=2$. When the condition \eqref{VC_cone} is satisfied for some $n_\perp$, there are also contributions from the harmonics $y^{(i)}_n=0$ with $n=\overline{-1,1}$, and $m=\{-2,0,2\}$ for the respective harmonic numbers $n$. In the case of negative $q$, the sign of $m$ is flipped provided all the other parameters remain unchanged. These selection rules agree with the general statement proved in \cite{BKL5} for the radiation of twisted photons produced by a charged particle moving along a straight line in a helical medium. They also agree with the selection rules obtained in the previous Sec. \ref{RadProbSec}. The comparison of the two approximations is presented in Fig. \ref{parax_anis_tw_plots} and shows a rather good agreement in the parameter space where the both approximations are valid.

\section{Conclusion}

Let us sum up the results. We developed QED in an anisotropic dispersive medium and constructed its particular case for a CLC plate of a finite width. The simplest model of a CLC was considered where the permittivity tensor has the form \eqref{permit_holec}. The medium with such a permittivity tensor is a particular case of a helical medium. It was proven in \cite{BKL5} that a charged particle moving along the axis of a helical medium produces the radiation which is a pure source of twisted photons at the radiation harmonics. Our explicit calculations in the present paper confirmed this general statement.

We constructed the quantum electromagnetic field operator in the presence of the CLC plate in the paraxial and small anisotropy approximations. In general, these approximations work in the different regions of parameters of the cholesteric and radiated photons and we found the parameter spaces where these approximations are applicable. Using these approximations, we obtained the mode functions, dispersion laws of photons, and the joining coefficients at the boundaries of the CLC plate. The interfaces of the plate were assumed to be normal to the axis of the cholesteric helix and the width of the plate was taken to be an integer number of the cholesteric pitch.

The explicit expression for a quantum electromagnetic field operator allows one to obtain the average number of plane-wave photons produced by a classical current \eqref{dP_plane}. Thereby we found the average number of plane-wave photons \eqref{aver_num_phot} radiated by a point charge traversing a cholesteric plate. As expected \cite{Ginzburg,BazylZhev,BelDmitOrl}, this radiation possesses harmonics with respect to the energy due to periodic structure of the CLC. Apart from the radiation probabilities, we derive the amplitudes of radiation of plane-wave photons by a charge moving along a straight line. We decomposed these amplitudes in terms of the twisted photons (see, e.g., \cite{JenSerprl,JenSerepj}) and obtained in this way the explicit expressions for the average number of twisted photons \eqref{dP_tw_par}, \eqref{dP_tw_anis} generated by a point charge traversing normally the CLC plate.

In the paraxial regime, we found the selection rules \eqref{rm_spectr}, \eqref{lm_spectr} for the projection $l$ of the orbital angular momentum of radiated twisted photons. The radiation of twisted photons with $l=\{-1,1\}$ dominates at the respective harmonics. The radiation of twisted photons with higher projection of orbital angular momentum is suppressed due to a peculiar form of the permittivity tensor of the CLC. In the parameter space where the perturbation theory with respect to the anisotropy works, the radiated twisted photons obey the selection rules \eqref{rm_spectr_m}, \eqref{lm_spectr_m} complying with the general selection rules proved in \cite{BKL5}. For the right-handed CLCs, $q>0$, the main contribution to radiation consists of the twisted photons with $m=2$ in the case when the VC cone is not formed and with $m=\{-2,0,2\}$ at the respective harmonics, otherwise. The sign flip of $q$ leads to the flip of signs of $m$, $s$, and $l$ provided all the other parameters are left intact.

As the explicit examples, the radiation of plane-wave and twisted photons produced in the optical range by the beams of electrons with the Lorentz factors $\ga=235$ and $\ga=500$ and the beam of uranium nuclei ${}^{238}$U${}^{92+}$ with $\ga=2$ were investigated. The particle beams are assumed to be Gaussian, collimated, and isoenergetic. The longitudinal dimension is taken $\s_3=150$ $\mu$m and the transverse size is $\s_\perp=1$ $\mu$m. The radiation of twisted photons from the particle beams is calculated by the use formulas presented in \cite{BKb,BKLb,BKL5}. In the case of the electrons with $\ga=500$ and the uranium nuclei, the CLC plate is assumed to be of the thickness $L=32$ $\mu$m, the CLC helix is supposed to be right-handed with the pitch $1.6$ $\mu$m, and so the number of periods $N_u=40$. For the energies of radiated photons we investigate, the parameters of the permittivity tensor are taken from \cite{LWGLW} and are given by $\e_\perp=2.22$ and $\e_\parallel=2.49$. As far the electrons with $\ga=235$ is concerned, the CLC plate is taken with the width $L=20$ $\mu$m, the CLC helix is right-handed with the pitch $1$ $\mu$m, and so the number of periods $N_u=40$. In the case of electrons with the Lorentz factors $\ga=500$ and $\ga=235$, the radiation produced in the CLC is not overlapped by the edge radiation produced at the plate interfaces only for sufficiently large angles to the surface normal, $n_\perp\gtrsim1/2$ (see Figs. \ref{parax_anis_plw_plots}, \ref{parax_anis_tw_plots}, \ref{anis_tw_plots}). Therefore, in order to create a pure source of twisted photons, a sufficiently narrow beam of electrons is needed since $k_\perp \s_\perp\ll1$ has to be satisfied for that \cite{BKb,BKLb}, where $k_\perp=k_0 n_\perp$ is the transverse photon momentum and $\s_\perp$ is the transverse size of the particle beam. In the case of uranium nuclei beam (see Figs. \ref{U_parax_plots}, \ref{parax_anis_plw_plots}), the radiation is concentrated at small $n_\perp$ and so the restrictions on the transverse beam size are not so stringent. Notice that even in the case of large $k_\perp\s_\perp$, the projection of the total angular momentum of radiated twisted photons per one photon is the same as for the one-particle radiation \cite{BKb,BKLb}.

In all the cases considered, the one-particle radiation at a given harmonic is a pure source of twisted photons with the definite projection of the total angular momentum, $m$ and, in the paraxial regime, with the definite projection of the orbital angular momentum, $l$. In the cases considered, the radiation of photons with the helicity $s=1$ is suppressed and the radiation escaping from the plate to the detector lies inside the both Cherenkov cones defined by the standard formulas for uniform media
\begin{equation}
    n^{\text{VC}_\perp}_\perp=\sqrt{\e_\perp-1/\be_3^2},\qquad n^{\text{VC}_\parallel}_\perp=\sqrt{\e_\parallel-1/\be_3^2}.
\end{equation}
For the parameters we consider, these two quantities are larger than $1$ for electrons and are larger than $0.94$ for ${}^{238}$U${}^{92+}$ with $\ga=2$. Therefore, as follows from the analysis given in Secs. \ref{ParaxAppr}, \ref{AnisotrAppr}, this radiation is affected by the anomalous Doppler effect \cite{GinzbThPhAstr,BKL8} and the main contribution comes from the harmonic with $n=-1$. Then the twisted photons at this harmonic possess  $m=-2$ and, in the paraxial regime, $l=-1$. This radiation can be amplified by the standard means using the coherent radiation from a bunch train of charged particles. Larger projections of the orbital angular momentum of twisted photons are produced in the coherent radiation from the helically microbunched beams \cite{HemMar12,HKDXMHR,BKLb} traversing the CLC plate.

\paragraph{Acknowledgments.}

This work was supported by the Russian Science Foundation (Project No. 17-72-20013).
%The reported study was funded by RFBR, project number 20-32-70023.

\appendix
\section{Unitarity relations}\label{UnitRels}

Consider the matrix differential equation of the second order
\begin{equation}\label{Schr_eqn}
    \big[\partial_zK(k_0;z)\partial_z+V(k_0;z)\big]u=0,
\end{equation}
where $K(k_0;z)$ and $V(k_0;z)$ are Hermitian $n\times n$ matrices for $k_0\in \mathbb{R}$, the elements of which are piecewise-continuous functions. We assume that
\begin{enumerate}
  \item $K(k_0;z)$ is nondegenerate and $K(k_0;z)|_{|z|\rightarrow\infty}$ are positive definite;
  \item $K(k_0;z)$ and $V(k_0;z)$ tend to constant values at $|z|\rightarrow\infty$ faster than $\exp(-c|z|)$ for any $c$ \cite{LandLifshQM.11};
  \item The eigenvalue problems,
\begin{equation}\label{eigen_prob}
    \big[-K(k_{0};z)\la^\pm_s(k_0)+V(k_{0};z)\big]_{z\rightarrow\pm\infty}f^\pm(s,k_0)=0,
\end{equation}
    where $s=\overline{1,n}$ numerates the solutions, possess the property that $\im\la^\pm_s(k_0)\neq0$ implies $\im k_0\neq0$.
\end{enumerate}
As for the Maxwell equations \eqref{Max_eqns2} describing the electromagnetic fields in the CLC plate of a finite width, these properties are satisfied and $\la^\pm_s(k_0)=k_0^2-k_\perp^2$.

If $\vf_1$ and $\vf_2$ are solutions of \eqref{Schr_eqn}, then their complex Wronskian does not depend on $z$:
\begin{equation}\label{Wronsk_const}
    W[\vf_1,\vf_2]:=\vf^\dag_1 K\partial_z\vf_2-\partial_z\vf^\dag_1 K\vf_2=const.
\end{equation}
Since
\begin{equation}
    W[\vf_1,\vf_2]=-W^*[\vf_2,\vf_1],
\end{equation}
the Wronskian defines an anti-Hermitian form on the solutions of Eq. \eqref{Schr_eqn}.

Let us standardly introduce the Jost functions
\begin{equation}\label{Jost_funcs}
    \big[\partial_zK(k_0;z)\partial_z+V(k_0;z)\big]F^\pm(s,k;z)=0,\qquad s=\overline{1,n},
\end{equation}
where
\begin{equation}\label{Jost_asympt}
    F^\pm(s,k;z)\underset{z\rightarrow\pm\infty}{\rightarrow}e^{ikz}f^\pm(s,k^2),\qquad \im k\geqslant0,
\end{equation}
and the vectors $f^\pm(s,k^2)$ obey Eqs. \eqref{eigen_prob} with
\begin{equation}
    k^2=\la^\pm_s(k_0).
\end{equation}
For real $k_0$, one can always put
\begin{equation}
    (f^\pm)^\dag(s,k_0)K_{z\rightarrow\pm\infty}f^\pm(s',k_0)=\de_{ss'}.
\end{equation}
Then, for real $k$ and $k_0$, it follows from the property \eqref{Wronsk_const} that
\begin{equation}\label{scalar_prod}
    W[F^\pm(s,k),F^\pm(s',k)]=2ik\de_{ss'},\qquad W[F^\pm(s,k),F^\pm(s',-k)]=0,
\end{equation}
where, for brevity, we do not write out the argument $z$ of the Jost functions.

In virtue of the property 2 and relations \eqref{scalar_prod}, the functions $F^+(s,\pm k)$ and $F^-(s,\pm k)$ constitute orthogonal bases in the $2n$-dimensional pseudo-Hilbert space of solutions of Eq. \eqref{Schr_eqn}. The two bases of solutions \eqref{Schr_eqn} are related as
\begin{equation}\label{2bases}
\begin{split}
    F^+(s,k)&=\al_{ss'}(k)F^-(s',k)+\be_{ss'}(k)F^-(s',-k),\\
    F^+(s,-k)&=\be_{ss'}(-k)F^-(s',k)+\al_{ss'}(-k)F^-(s',-k),
\end{split}
\end{equation}
where the summation is understood with respect to repeated indices. Using \eqref{scalar_prod} and \eqref{2bases}, one infers that
\begin{equation}\label{unitar1}
    \al(k)\al^\dag(k)-\be(k)\be^\dag(k)=1,\qquad \al(-k)\be^\dag(k)-\be(-k)\al^\dag(k)=0,
\end{equation}
where
\begin{equation}
    \left[
      \begin{array}{cc}
        \al(k) & \be(k) \\
        \be(-k) & \al(-k) \\
      \end{array}
    \right]\in U(n,n).
\end{equation}

The relations \eqref{unitar1} are equivalent to the unitarity relations of the respective scattering matrix. Introduce the basis of solutions of Eq. \eqref{Schr_eqn} of the form
\begin{equation}
\begin{split}
    \vf_1(s,k)&=t^{(1)}_{ss'}(k)F^+(s',k)=F^-(s,k)+r^{(1)}_{ss'}(k)F^-(s',-k),\\
    \vf_2(s,k)&=t^{(2)}_{ss'}(k)F^-(s',-k)=F^+(s,-k)+r^{(2)}_{ss'}(k)F^+(s',k).
\end{split}
\end{equation}
Then
\begin{equation}
\begin{gathered}
    t_{(1)}(k)=\al^{-1}(k),\qquad r_{(1)}(k)=\al^{-1}(k)\be(k),\qquad r_{(2)}(k)=-\be(-k)\al^{-1}(k),\\
    t_{(2)}^\dag(-k)=t_{(1)}(k),\qquad r_{(1,2)}^\dag(-k)=r_{(1,2)}(k),
\end{gathered}
\end{equation}
and the matrix
\begin{equation}
    S(k):=
    \left[
      \begin{array}{cc}
        t_{(1)}(k) & r_{(1)}(k) \\
        r_{(2)}(k) & t_{(2)}(k) \\
      \end{array}
    \right]
\end{equation}
is unitary.

The definition of the Jost functions and the relations \eqref{2bases} are valid for complex $k$ as well. It follows from \eqref{Jost_asympt} and \eqref{2bases} that the spectrum of bound states is determined by the equation
\begin{equation}\label{resonanses}
    \det\al(k)=0,\qquad\im k>0,
\end{equation}
This condition provides the existence of the nontrivial linear combination of the Jost functions $F^+(s,k)$ that exponentially tends to zero for $|z|\rightarrow\infty$. The property 3 mentioned above and the absence of singularities of $H^{-1}(k_0)$ out of the real axis for transparent media (see Sec. \ref{Gener_Form}) imply that the spectrum of bound states lies only at $k=i|k|$. The branch points of $\det\al(k)$ for $k\geqslant0$ correspond to the boundaries of the continuous spectrum. The zeros of $\det\al(k)$ on the unphysical sheet of the complex $k$-plane, $\im k<0$, specify the resonances and virtual states. The virtual (anti-bound) states are located at $\re k=0$ on the unphysical sheet.

\section{Joining the mode functions}\label{JoinModFun}
\subsection{Paraxial approximation}\label{ParAx_App}

The coefficients of the Fourier series \eqref{mode_gen} for the mode $1$ are written as
\begin{equation}\label{a1_kp}
\begin{split}
    a^{(1)}_+(p_3+2q)&=-\frac{k_\perp^2\de\e}{64q} \frac{q+v_++\bar{k}_0w+\de\e\bar{k}_0^2/(4q)}{q^2+qv_+ +\bar{k}_0w(3q+v_+)/2+(1+\de\e/2)\bar{k}_0^2/2},\\
    a^{(1)}_-(p_3)&=-\frac{5k_\perp^2}{16\bar{k}_0^2}\frac{(q+3v_+/5+\bar{k}_0w/5)(q+v_++\bar{k}_0w+\de\e\bar{k}_0^2/(4q))}{q^2+qv_+ +\bar{k}_0w(3q+v_+)/2+(1+\de\e/2)\bar{k}_0^2/2},\\
    a^{(1)}_+(p_3-2q)&=-\frac{5k_\perp^2}{16\bar{k}_0^2}\frac{(q-3v_+/5+\bar{k}_0w/5)(q+v_++\bar{k}_0w+\de\e\bar{k}_0^2/(4q))}{q^2-qv_+ +\bar{k}_0w(3q-v_+)/2+(1+\de\e/2)\bar{k}_0^2/2},\\
    a^{(1)}_-(p_3-2q)&=a^{(1)}_{0-}(p_3-2q)+\frac{2k_\perp^2q^2}{\de\e\bar{k}_0^4}\Big[1+v_+^{-1}+\frac{\bar{k}_0}{wv_+} \Big(1+\frac{\de\e}{4}\Big(3+v_+ +\frac{\bar{k}_0w}{q}\Big)+\frac{\de\e^2\bar{k}_0^2}{16q^2}\Big)\Big],\\
    a^{(1)}_-(p_3-4q)&=-\frac{k_\perp^2\de\e}{64q} \frac{q+v_++\bar{k}_0w+\de\e\bar{k}_0^2/(4q)}{q^2-qv_+ +\bar{k}_0w(3q-v_+)/2+(1+\de\e/2)\bar{k}_0^2/2},
\end{split}
\end{equation}
up to the terms of the order $k_\perp^2$. The analogous coefficients for the mode $2$ have the form
\begin{equation}\label{a2_kp}
\begin{split}
    a^{(2)}_+(p_3+2q)&=-\frac{k_\perp^2\de\e}{64q} \frac{q+v_--\bar{k}_0w+\de\e\bar{k}_0^2/(4q)}{q^2+qv_- -\bar{k}_0w(3q+v_-)/2+(1+\de\e/2)\bar{k}_0^2/2},\\
    a^{(2)}_-(p_3)&=-\frac{5k_\perp^2}{16\bar{k}_0^2}\frac{(q+3v_-/5-\bar{k}_0w/5)(q+v_--\bar{k}_0w+\de\e\bar{k}_0^2/(4q))}{q^2+qv_- -\bar{k}_0w(3q+v_-)/2+(1+\de\e/2)\bar{k}_0^2/2},\\
    a^{(2)}_+(p_3-2q)&=-\frac{5k_\perp^2}{16\bar{k}_0^2}\frac{(q-3v_-/5-\bar{k}_0w/5)(q+v_--\bar{k}_0w+\de\e\bar{k}_0^2/(4q))}{q^2-qv_- -\bar{k}_0w(3q-v_-)/2+(1+\de\e/2)\bar{k}_0^2/2},\\
    a^{(2)}_-(p_3-2q)&=a^{(2)}_{0-}(p_3-2q)+\frac{2k_\perp^2q^2}{\de\e\bar{k}_0^4}\Big[1+v_-^{-1}-\frac{\bar{k}_0}{wv_-} \Big(1+\frac{\de\e}{4}\Big(3+v_- -\frac{\bar{k}_0w}{q}\Big)+\frac{\de\e^2\bar{k}_0^2}{16q^2}\Big)\Big],\\
    a^{(2)}_-(p_3-4q)&=-\frac{k_\perp^2\de\e}{64q} \frac{q+v_--\bar{k}_0w+\de\e\bar{k}_0^2/(4q)}{q^2-qv_- -\bar{k}_0w(3q-v_-)/2+(1+\de\e/2)\bar{k}_0^2/2},
\end{split}
\end{equation}
up to the terms of the order $k_\perp^2$. It is seen from \eqref{disper_parax1}, \eqref{disper_parax2}, \eqref{a1_kp}, and \eqref{a2_kp} that the mode $2$ can be obtained from the mode $1$ with the aid of the replacements $w\rightarrow-w$ and $\ups_+\rightarrow\ups_-$, i.e., in fact, by changing the sign in the definition of the square root in $w$.

Introduce the notation
\begin{equation}
    a_{1,2}:=a^{(1,2)}_{0-}(p_3-2q),\qquad p_{1,2}:=p^{(1,2)}_0,
\end{equation}
and
\begin{equation}
\begin{split}
    d_n:&=[p_1(a_1-1)(a_2+1) -p_2(a_1+1)(a_2-1)-2a_1+2a_2]\times\\
    &\times [p_1(a_1+1)(a_2-1) -p_2(a_1-1)(a_2+1)+2a_1-2a_2],
\end{split}
\end{equation}
and
\begin{equation}
\begin{split}
    \zeta^1_\pm:&=(\pm k_0-p_2+2q)(p_1-p_2)a_1a_2^2-(\pm k_0+p_2)(p_1+p_2-2q)a_1+2(p_2-q)(\pm k_0+p_1)a_2,\\
    \zeta^2_\pm:&=(\pm k_0-p_1+2q)(p_2-p_1)a_2a_1^2-(\pm k_0+p_1)(p_2+p_1-2q)a_2+2(p_1-q)(\pm k_0+p_2)a_1,\\
    \s^1_\pm:&=(\pm k_0+p_2+2q)(p_1+p_2-2q)a_2^2 -2(p_2-q) (\pm k_0+p_1-2q)a_1a_2 -(\pm k_0-p_2)(p_1-p_2),\\
    \s^2_\pm:&=(\pm k_0+p_1+2q)(p_2+p_1-2q)a_1^2 -2(p_1-q) (\pm k_0+p_2-2q)a_2a_1 -(\pm k_0-p_1)(p_2-p_1).
\end{split}
\end{equation}
Then, in the case $k_\perp=0$, the coefficients $\bar{r}_i$ and $\bar{l}_i$ of the linear combination defined in \eqref{bar_r_bar_l} read for $s=1$ as
\begin{equation}\label{r_l_expl}
    \bar{r}_i=\sqrt{2}d_n^{-1}\zeta^i_+,\qquad \bar{l}_i=-\sqrt{2}d_n^{-1}\s^i_-.
\end{equation}
If $s=-1$, then
\begin{equation}
    \bar{r}_i=-\sqrt{2}d_n^{-1}\s^i_+,\qquad \bar{l}_i=\sqrt{2}d_n^{-1}\zeta^i_-,
\end{equation}
i.e., the sign flip of the helicity gives rise to the change $\zeta\leftrightarrow-\s$ in \eqref{r_l_expl}.

It is not difficult to find the explicit expressions for the coefficients $a_l$ of the linear combination in the leading order in $k_\perp$. In the limit $k_\perp=0$, we have
\begin{equation}
\begin{split}
    a_l=&\,\frac{1}{2\sqrt{2}k_0}\diag(e^{ik_3L-i(s-1)\vf},e^{ik_3L-i(s+1)\vf},-e^{-ik_3L-i(s+1)\vf},-e^{-ik_3L-i(s-1)\vf})\times\\
    &\times\left[
      \begin{array}{cccc}
        (k_0+p_1-2q)a_1 & (k_0+p_2-2q)a_2 & k_0-p_1 & k_0-p_2 \\
        k_0+p_1 & k_0+p_2 & (k_0-p_1+2q)a_1 & (k_0-p_2+2q)a_2 \\
        k_0-p_1 & k_0-p_2 & (k_0+p_1-2q)a_1 & (k_0+p_2-2q)a_2 \\
        (k_0-p_1+2q)a_1 & (k_0-p_2+2q)a_2 & k_0+p_1 & k_0+p_2 \\
      \end{array}
    \right]\times\\
    &\times\diag(e^{-ip_3^{(1)}L},e^{-ip_3^{(2)}L},e^{ip_3^{(1)}L},e^{ip_3^{(2)}L})
    \left[
      \begin{array}{c}
        \bar{r}_1 \\
        \bar{r}_2 \\
        \bar{l}_1 \\
        \bar{l}_2 \\
      \end{array}
    \right].
\end{split}
\end{equation}
In this limit, the normalization constant \eqref{norm_const0} becomes
\begin{equation}
    |c|^{-2}=\nu_0+\big[\nu_1 e^{-i(p^{(1)}_3+p^{(2)}_3)L} +\nu_2 e^{-2ip^{(1)}_3L} +\nu_3 e^{-2ip^{(2)}_3L} +\nu_4 e^{i(p^{(1)}_3-p^{(2)}_3)L} +c.c.\big],
\end{equation}
where
\begin{equation}
\begin{split}
    \nu_0=&\,\frac{1}{4k_0^2d_n^2} \sum_{i=1}^2\Big\{[(k_0+p_i)^2+(k_0+p_i-2q)^2a_i^2](\zeta_+^i)^2 +[(k_0-p_i)^2+(k_0-p_i+2q)^2a_i^2](\s^i_-)^2 \Big\},\\
    \nu_1=&\,-\frac{1}{4k_0^2d_n^2}\Big\{[(k_0-p_2)(k_0+p_1-2q)a_1 +(k_0+p_1)(k_0-p_2+2q)a_2]\zeta_+^1\s^2_-+\\
    &+[(k_0-p_1)(k_0+p_2-2q)a_2 +(k_0+p_2)(k_0-p_1+2q)a_1]\zeta_+^2\s^1_- \Big\},\\
    \nu_2=&\,-\frac{a_1}{2k_0^2d_n^2} (k_0^2-p_1(p_1-2q))\zeta^1_+\s^1_-,\\
    \nu_3=&\,-\frac{a_2}{2k_0^2d_n^2} (k_0^2-p_2(p_2-2q))\zeta^2_+\s^2_-,\\
    \nu_4=&\,\frac{1}{4k_0^2d_n^2}\Big\{[(k_0+p_1)(k_0+p_2) +(k_0+p_1-2q)(k_0+p_2-2q)a_1a_2 ]\zeta_+^1\zeta_+^2+\\
    &+[(k_0-p_1)(k_0-p_2) +(k_0-p_1+2q)(k_0-p_2+2q)a_1a_2 ]\s_-^1\s_-^2 \Big\},
\end{split}
\end{equation}
for $s=1$. The normalization constant for $s=-1$ is obtained by the replacement $\zeta\leftrightarrow\s$ in the above expressions for $\nu_k$.

\subsection{Perturbation theory in anisotropy}\label{de_App}

The coefficients of the Fourier series \eqref{mode_gen} for the mode $1$ are
\begin{equation}\label{a1_de}
\begin{split}
    a^{(1)}_+(p_3)&=1,\\
    a^{(1)}_-(p_3)&=1-\frac{\de\e\bar{k}_3 \bar{k}_0^2}{8qk_\perp^2}\frac{2\bar{k}_3^2+k_\perp^2}{\bar{k}_3^2-q^2},\\
    a^{(1)}_+(p_3+2q)&=\frac{\de\e}{16q}\frac{2\bar{k}_3^2+k_\perp^2}{\bar{k}_3+q},\\
    a^{(1)}_-(p_3+2q)&=-\frac{\de\e}{16q}\frac{k_\perp^2}{\bar{k}_3+q},\\
    a^{(1)}_+(p_3-2q)&=\frac{\de\e}{16q}\frac{k_\perp^2}{\bar{k}_3-q},\\
    a^{(1)}_-(p_3-2q)&=-\frac{\de\e}{16q}\frac{2\bar{k}_3^2+k_\perp^2}{\bar{k}_3-q},
\end{split}
\end{equation}
up to the terms of the order $\de\e$. As for the mode $2$, we have
\begin{equation}\label{a2_de}
\begin{split}
    a^{(2)}_+(p_3)&=1,\\
    a^{(2)}_-(p_3)&=-1-\frac{\de\e\bar{k}_3^3}{8qk_\perp^2}\frac{2\bar{k}_3^2+k_\perp^2}{\bar{k}_3^2-q^2},\\
    a^{(2)}_+(p_3+2q)&=-\frac{\de\e}{16q}\frac{2\bar{k}_3^2+k_\perp^2}{\bar{k}_3+q},\\
    a^{(2)}_-(p_3+2q)&=\frac{\de\e}{16q}\frac{k_\perp^2}{\bar{k}_3+q},\\
    a^{(2)}_+(p_3-2q)&=\frac{\de\e}{16q}\frac{k_\perp^2}{\bar{k}_3-q},\\
    a^{(2)}_-(p_3-2q)&=-\frac{\de\e}{16q}\frac{2\bar{k}_3^2+k_\perp^2}{\bar{k}_3-q},
\end{split}
\end{equation}
up to the terms of the order $\de\e$.

At zeroth order in $\de\e$, the coefficients $a_{ch}$, $a_l$ of the linear combination of the solutions of the Maxwell equations \eqref{Max_eqns3} are as follows
\begin{equation}\label{ach_al_0}
\begin{split}
    a_{ch}^T=&\,\frac{1}{2\sqrt{2}} \Big[\vs_{1}\frac{\e_\perp n_3+\bar{n}_3}{\e_\perp} , -s\vs_{2}\frac{n_3+\bar{n}_3}{\bar{n}_3},\frac{\e_\perp n_3-\bar{n}_3}{\vs_{1}\e_\perp }, -s\frac{n_3-\bar{n}_3}{\vs_{2}\bar{n}_3} \Big],\\
    a_l^T=&\,\frac{1}{8} \Big[\frac{(\e_\perp n_3+\bar{n}_3)^2}{\vk_{1}\e_\perp n_3\bar{n}_3} -\vk_{1}\frac{(\e_\perp n_3-\bar{n}_3)^2}{\e_\perp n_3\bar{n}_3} +s\frac{(n_3+\bar{n}_3)^2}{\vk_{2} n_3\bar{n}_3} -s\vk_{2}\frac{(n_3-\bar{n}_3)^2}{ n_3\bar{n}_3},\\
    &\frac{(\e_\perp n_3+\bar{n}_3)^2}{\vk_{1}\e_\perp n_3\bar{n}_3} -\vk_{1}\frac{(\e_\perp n_3-\bar{n}_3)^2}{\e_\perp n_3\bar{n}_3} -s\frac{(n_3+\bar{n}_3)^2}{\vk_{2} n_3\bar{n}_3} +s\vk_{2}\frac{(n_3-\bar{n}_3)^2}{ n_3\bar{n}_3},\\
    & (\vk^{-1}_{1}-\vk_{1}) \Big(\frac{\e_\perp n_3}{\bar{n}_3} -\frac{\bar{n}_3}{\e_\perp n_3} \Big) +s(\vk^{-1}_{2}-\vk_{2}) \Big(\frac{n_3}{\bar{n}_3} -\frac{\bar{n}_3}{n_3} \Big),\\
    & (\vk^{-1}_{1}-\vk_{1}) \Big(\frac{\e_\perp n_3}{\bar{n}_3} -\frac{\bar{n}_3}{\e_\perp n_3} \Big) -s(\vk^{-1}_{2}-\vk_{2}) \Big(\frac{n_3}{\bar{n}_3} -\frac{\bar{n}_3}{n_3}\Big) \Big]\times\\
    &\times\diag(e^{ik_3L},e^{ik_3L},e^{-ik_3L},e^{-ik_3L}),
\end{split}
\end{equation}
where $\bar{n}_3:=\bar{k}_3/k_0$ and the terms proportional to $\de\e$ are retained in the exponents
\begin{equation}\label{vsi}
    \vs_{i}:=e^{ip^{(i)}_3\vf/q},\qquad \vk_{i}:=e^{ip_3^{(i)}L}.
\end{equation}
For $\de\e=0$, the normalization constant takes the form
\begin{equation}\label{c_0}
    |c|^{-2}=\frac{1}{16}\Big[12 +\frac{(\e_\perp^2+1)(\e_\perp^2 n_3^4+\bar{n}_3^4)}{\e_\perp^2n_3^2\bar{n}_3^2} -\cos^2(p_3^{(1)}L)\frac{(\e_\perp^2n_3^2-\bar{n}_3^2)^2}{\e_\perp^2 n_3^2\bar{n}_3^2} -\cos^2(p_3^{(2)}L)\frac{(n_3^2-\bar{n}_3^2)^2}{n_3^2\bar{n}_3^2}\Big].
\end{equation}

\end{document}